\documentclass[12pt]{article}
\usepackage{amsmath}
\usepackage{amsfonts}
\usepackage{graphicx,psfrag,epsf}
\usepackage{enumerate}
\usepackage{amsthm}
\usepackage{color}
\usepackage{booktabs}
\newcommand{\blind}{0}

\newtheorem{theorem}{Theorem}

\begin{document}

\bibliographystyle{abbrvnat}

\def\spacingset#1{\renewcommand{\baselinestretch}%
{#1}\small\normalsize} \spacingset{1}

%%%%%%%%%%%%%%%%%%%%%%%%%%%%%%%%%%%%%%%%%%%%%%%%%%%%%%%%%%%%%%%%%%%%%%%%%%%%%%

\if0\blind
{
  \title{\bf Density Estimation on a Network}
  \author{Yang Liu\thanks{
    The authors gratefully acknowledge}\hspace{.2cm}\\
    Department of Statistics and Data Science, Cornell University\\
    and \\
    David Ruppert \\
    Department of Statistics and Data Science and \\ School of Operations Research and Information Engineering, Cornell University}
  \maketitle
} \fi

\if1\blind
{
  \bigskip
  \bigskip
  \bigskip
  \begin{center}
    {\LARGE\bf Title}
\end{center}
  \medskip
} \fi

\bigskip
\begin{abstract}
This paper develops a novel approach to density estimation on a network. We formulate nonparametric density estimation on a network as a nonparametric regression problem by binning. Nonparametric regression using local polynomial kernel-weighted least squares have been studied rigorously \cite{ruppert1994multivariate}, and its asymptotic properties make it superior to kernel estimators such as the Nadaraya-Watson estimator. When applied to a network, the best estimator near a vertex depends on the amount of smoothness at the vertex. Often, there are no compelling reasons to assume that a density will be continuous or discontinuous at a vertex, hence a data driven approach is proposed. To estimate the density in a neighborhood of a vertex, we propose a two-step procedure. The first step of this pretest estimator fits a separate local polynomial regression on each edge using data only on that edge, and then tests for equality of the estimates at the vertex. If the null hypothesis is not rejected, then the second step re-estimates the regression function in a small neighborhood of the vertex, subject to a joint equality constraint. Since the derivative of the density may be discontinuous at the vertex, we propose a piecewise polynomial local regression estimate to model the change in slope.  We study in detail the special case of local piecewise linear regression and derive the leading bias and variance terms using weighted least squares theory. We show that the proposed approach will remove the bias near a vertex that has been noted for existing methods, which typically do not allow for discontinuity at vertices. For a fixed network, the proposed method scales sub-linearly with sample size and it can be extended to regression and varying coefficient models on a network. We demonstrate the workings of the proposed model by simulation studies and apply it to a dendrite network data set.
\end{abstract}

\noindent%
{\it Keywords:}  Asymptotic bias and variance, discontinuous density, kernel density estimation, local piecewise linear estimation, pretest estimation

\spacingset{1.45}

\section{Introduction}
\begin{figure}[h]
\centering
\includegraphics[scale=0.8]{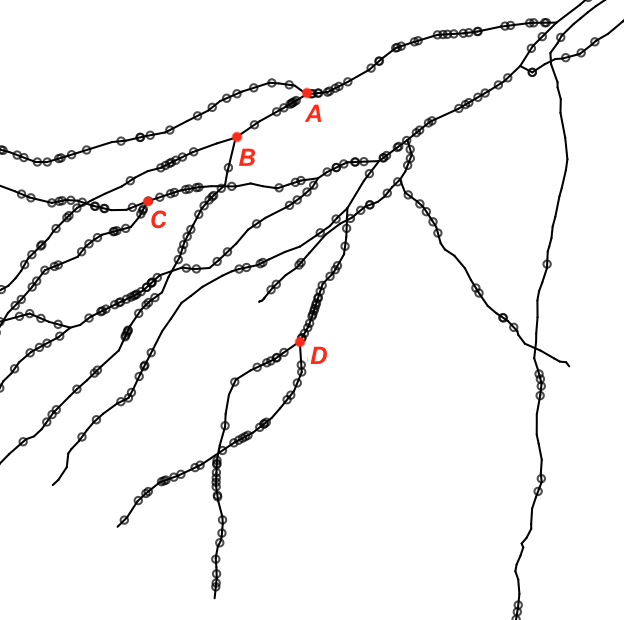}
\caption{Dendritic spines on a branch of the dendrite network of a neuron.}
\label{fig:dendriticSplines}
\end{figure}

In this paper, we analyze spatial patterns of points that lie on a network. There are many types of events that can occur along networks. The lines or curves that form the network may be roads, rivers, subway lines, spider webs, blood vessels, or nerve fibers. The events may be vehicles, street crimes, car accidents, retail stores, or neuroanatomical features such as dendritic spines. For a recent summary of applications that involve the modelling of point patterns on a network of lines, see \cite{baddeley2015spatial}. In order to carry out analyses of those events, researchers need a range of specific techniques.\\
\\
One of the frequently demanded tasks is density estimation on a network, but few statistical methods had been developed to address this need until recently. A natural first attempt at analyzing such data is to take the kernel density estimate (KDE) on the one-dimensional real line 
\begin{align}
\label{kde}
    \hat{f}(x)=\frac{1}{N}\sum^N_{i=1}K_h(x_i-x),
\end{align}
and apply it directly to network data by defining $\vert x_i-x\vert$ as the network distance, where $x$ is any location on the network, and $x_1,\dots,x_N$ are the observed data locations. $K_h(\cdot)=K(\cdot/h)/h$, where $K$ is a kernel function, and $h$ is the bandwidth. The network distance is defined as the shortest path distance between two points on a network. However, under this approach, estimate \eqref{kde} does not conserve mass. This happens because that the induced kernel $K_h(\cdot)$ is not a probability density on the network at points within the $h$-neighborhood of a vertex, and so \eqref{kde} is not a probability density. As a result, it will overestimate the true density. \cite{okabe2012spatial} summarized widely used methods for density estimation on a network. Many published papers, such as \cite{xie2008kernel}, mentioned by Okabe et al.\ computed a kernel density estimate on a network, but most have used \eqref{kde}.\\
\\
Although there is currently no consensus view on how to perform density estimation on a network, the two most popular heuristic techniques, namely the equal-split discontinuous kernel (ESDK) estimator and the equal-split continuous kernel (ESCK) estimator (\cite{okabe2009kernel}, \cite{okabe2012spatial}, \cite{sugihara2010simple}) seem to produce reasonable results in applications. However their methods suffer from a lack of theoretical grounding, and their computational cost is high (see Section 7.2). Furthermore, ESCK does not allow for discontinuity at vertices, which is found in many applications, such as traffic network, and the example in Figure 1 and Section 7.4. Although ESDK produces discontinuous density at vertices, the estimation in the small neighborhood of vertices always uses data from all edges, which can lead to large bias. This is demonstrated in Case I of the simulation study in Section 7.1. \\
\\
More recently \cite{mcswiggan2017kernel} proposed a diffusion density estimator (DE) on networks using the connection between kernel smoothing and diffusion \cite{botev2010kernel}. DE is based on numerical solution of the heat equation, and it is not only faster than ESDK and ESCK, but also provides a sound statistical rationale and helps establish theoretical properties. The estimate is asymptotically unbiased with rate $h^2$ away from the boundary region, where $h$ is the bandwidth.  At a terminal vertex, it has the standard boundary rate $h$ of \mbox{KDE}. Moreover, similar to ESCK, DE does not allow for discontinuity at vertices, which consequently, can lead to large bias.\\
\\
In this paper, we take a different approach. We propose local piecewise linear regression (LPLR) on a network, and establish theoretical properties for the proposed method. The piecewise linear structure is needed to model the discontinuity in the regression function or the first derivative of the regression function at a vertex. By binning, we form a histogram of observations (bin centers and counts) on the network, which is then smoothed by local polynomial regression (LPR). The advantage of local polynomial estimation is reduction of the order of the bias, especially when the evaluation point is near or on the boundary vertex, and, as a practical matter, the amount of computation is reduced. To accommodate densities with discontinuous derivatives at vertices, we introduce piecewise local regression. As far as we know, local polynomial regression on networks has not be proposed and investigated and piecewise local linear regression is also new. We note that the proposed method assumes that the data comes from a Poisson process on a network, but McSwiggan et al is applicable to more general point processes. However we want to point out that our approach has better order of the bias near vertices, and it only imposes continuity at vertices when there is evidence suggested by the data that the density is continuous there. Our approach is not limited to density estimation as it can be extended to handle other types of analysis of network data, such as regression analysis. The implementation of the proposed method is fast, the computation time scales only with data sub-linearly.  \\
\\
To motivate our research and illustrate its contribution, we introduce the dendrite data collected by the Kosik Lab, UC Santa Barbara, and first analyzed by \cite{baddeley2014multitype} and \cite{jammalamadaka2013statistical}. In this example, the events are dendritic splines, which are of clinical importance. Cognitive disorders such as ADHD and autism may result from abnormalities in dendritic spines, especially the number of spines and their maturity \cite{penzes2011dendritic}. The events on the network are the locations of 566 spines observed on one branch of the dendritic tree of a rat neuron, as shown in Figure~\ref{fig:dendriticSplines}. The density of the spines shows clear discontinuity at vertices A, C and D; see Figures~\ref{est_a_new}, \ref{est_b_new}, \ref{est_c_new} and \ref{est_d_new}. We will need a density estimation method that allows for multiple levels of smoothness at vertices -- discontinuous, continuous with discontinuous derivative, and continuous with continuous derivative. We show that our proposed method will remove the bias that has been noted for existing methods such as \cite{okabe2012spatial} and \cite{mcswiggan2017kernel} due to their inflexibility at vertices. KDE, ESDK, ESCK and DE's inflexibility arises because the smoothness of the density function at a vertex needs to be decided before choosing an estimator: if discontinuity is desired at a vertex, KDE is applied to each edge, otherwise ESDK, ESCK or DE should be applied. In contrast, we propose a data-driven approach and determine the smoothness of the density function at a vertex by statistical testing. In fact, estimation at vertices is a major contribution of this paper.\\
\\
The paper is organized as follows. Section~\ref{sec:preliminaries} gives basic definitions. In Section~\ref{sec:LPR}, a binned local polynomial regression estimator is constructed individually on each edge, and asymptotic properties of the estimator are established. As the evaluation point approaches the vertex from each edge, we obtain limits of the regression functions. In Section \ref{sec:test}, we construct an asymptotic test for the equality of those limits. If the null hypothesis is not rejected, then the regression function is re-estimated for evaluation points that are within the $h$-neighborhood of a vertex by binned local piecewise polynomial regression using data from all neighboring edges. These estimators are constructed and their asymptotic properties are studied in Section~\ref{sec:estEqual}. In Section~\ref{sec:practical}, we discuss networks with loops and related computational issues when the bandwidth is large. In Section~\ref{sec:implement}, we demonstrate the workings of the proposed method with simulated data and a real example, and discuss and analyze practical issues. We end with a discussion in Section \ref{sec:discuss}.

\section{Preliminaries}
\label{sec:preliminaries}
In this section, we introduce basic definitions and formulate density estimation as a local linear regression problem.
\subsection{Network and density function on a network}
Let $v_1, v_2\in\mathbb{R}^d$ be vertices and define an edge connecting $v_1$ and $v_2$ as a curve $e:[0,1]\to\mathbb{R}^d$ where $e(0) = v_1$ and $e(1) = v_2$. We call $L=\cup^J_{i=j}e_j$, $0<J<\infty$ a network in $\mathbb{R}^d$. A path between two points $x,y\in L$ is a sequence $\pi=(x,v_1,\dots,v_n,y)$, where $v_1,\dots,v_n$ are vertices, such that $e_j=[v_j,v_{j+1}]$ are edges, and $x\in e_0$, $y\in e_n$. The path length is Length$(\pi)=\int_x^{v_1}|e_0'(t)|dt+\sum_{j=1}^{n-1} \int^{v_{j+1}}_{v_j}|e_j'(t)|dt+\int_{v_n}^y|e_n'(t)|dt$, where integration is with respect to arc length. In the case where there are multiple paths between two points, we will use the shortest path to define length. If there no paths from $x$ to $y$, we define the network distance between $x$ and $y$ to be infinity. Often, the network is embedded in a two-dimensional space, such as a street network (if there are no overpasses). However all of our results apply easily to a network of curves embedded in a higher dimensional space, such as the dendrites which are embedded in $\mathbb{R}^3$.  In the theory section (Section \ref{sec:LPR}, \ref{sec:test} and \ref{sec:estEqual}), we will study in deal a simple network, where there is one vertex connecting $J$ edges. We discuss application to  complex networks in Section \ref{sec:practical} and \ref{sec:implement}.\\
\\
Our objective is to construct an estimate of an unobservable  underlying probability density function $f(x)$ of event locations based on observed data $x_1,\cdots, x_n\in L$. The density satisfies $f(x)\ge0$ for all $x\in L$, and $\int_Lf(x)dx=\sum_{j=1}^J\int_{e_j}f(x)dx=1$. Stated differently, let $\boldsymbol{x}=\{x_1,\dots,x_n\}$ be a realization of a Poisson process $\boldsymbol{X}$ on the network. We want to estimate the intensity function $\lambda(x)$ of $\boldsymbol{X}$. The intensity function $\lambda(x)$ is the expected number of random points per unit length of network in the small neighborhood of $x$. The estimation of $f$ and of $\lambda$ are equivalent because, for fixed $N$, $\lambda(x)=Nf(x)$ for all $x\in L$.
\subsection{Problem formulation}
In this paper, we consider density estimation via local polynomial regression. This is achieved by way of binning. Here we use the ``simple binning'' discussed in \cite{hall1996accuracy}. For $i=1,\dots,n$, $x_i$  are the bin centers, $c_i$ are bin counts, $y_i$ are bin heights, and $\omega$ is bin width. Letting $N$ be the total number of observations on the linear network, we have $y_i={c_i}/{N\omega}$. These rectangles form a histogram with total area $1$, since the area of the $i$th bin is $c_i/N$ and $\sum_{i=1}^n c_i/N=1$. For a chosen $\omega$, we consider the regression model $y_i=m(x_i)+\epsilon(x_i)$. We have
\begin{align}
    E(y_i|x_i)=E\left(\frac{c_i}{N\omega}\right)=\frac{1}{\omega}E(\hat{p}_i)=\frac{1}{\omega}p_i \approx m(x_i),\notag
\end{align}
where $p_i$ is the expected proportion of points in bin $i$, and $m(\cdot)$ is the regression function. We note that this assumption would still hold for more general point processes. For the variance, $\hat{p}_i=c_i/N$ is the sample proportion so 
\begin{align}
    Var(y_i|x_i)=Var\left(\frac{c_i}{N\omega}\right)=\frac{1}{\omega^2}Var(\hat{p}_i)=\frac{p_i-p_i^2}{N\omega^2}\approx\frac{1}{N\omega}m(x_i)+\frac{1}{N}m(x_i)^2.\notag
\end{align}
The variance expression depends on the Poisson process assumption. The approximate equality symbol is due to binning induces a bias, but the bias can be made negligible by a small bin width. Hall and Wand studied the accuracy of binned kernel density estimators \cite{hall1996accuracy}, and their results show that, with the commonly-used Epanechnikov kernel, we require $\omega$ to go to zero faster than $h$ the bandwidth, if binning is not to have a significant effect on the bias of the estimator. Therefore, the error term $\epsilon(x_i)$ has approximately mean zero and variance $m(x_i)/N\omega+m(x_i)^2/N$.\\
\\
Assuming that $\omega$ is sufficiently small, we can smooth the histogram by local polynomial regression using $(x_i,y_i)$, $i=1,\dots,n$, as data. We propose a ``pretest'' estimation procedure that consists of the following steps:
\begin{enumerate}
\item Local polynomial regression on each edge individually.
\item Test joint equality of intercepts at the vertex.
\item If joint equality is not rejected, then the regression function is re-estimated by local piecewise polynomial regression using data from all neighboring edges for evaluation points that are within the $h$-neighborhood of a vertex.
\end{enumerate}
An asymptotic test for joint equality of slopes at the vertex can be constructed similarly.

\section{Local polynomial regression}
\label{sec:LPR}
First, we consider each edge individually. This is equivalent to a fixed equally-spaced design model, where the $x$-variables are the bin centers $x_1,\dots,x_n$, and the $y$-variables are the bin heights $y_1,\dots,y_n$. We want to estimate the regression function $m(x)=E(Y|X=x)$. Using a Taylor expansion, we can approximate $m(x)$, where $x$ is close to a point $x_0$, by a $p$ degree polynomial:
\begin{align}
    m(x)&\approx m(x_0)+m^{(1)}(x_0)(x-x_0)+\frac{m^{(2)(x_0)}}{2!}(x-x_0)^2+\dots+\frac{m^{(p)}(x_0)}{p!}(x-x_0)^p\notag\\
    &=m(x_0)+\beta_1(x-x_0)+\beta_2(x-x_0)^2+\dots+\beta_p(x-x_0)^2,\notag
\end{align}
provided that all the required derivatives exist. The local polynomial regression estimator minimizes with respect to $\beta_0,\beta_1,\dots,\beta_p$ the function
\begin{align}
\label{mini}
\sum^n_{i=1}\big\{y_i-\beta_0-\beta_1(x_i-x_0)-\dots-\beta_p(x_i-x_0)^p\big\}^2K_h\left({x_i-x_0}\right),
\end{align}
where $K_h\left({x_i-x_0}\right)=K(\frac{x_i-x_0}{h})/h$. Let $\hat{\boldsymbol{\beta}}=\{\hat{\beta_0},\hat{\beta_1},\dots,\hat{\beta_p}\}$ denote the  minimizer of \eqref{mini}. Then $\hat{\beta_0}$ estimates $m(x)$ and $s!\hat{\beta_s}$ estimates $m^{(s)}(x)$, the $s$th derivative of $m(x)$, for $s=1,\dots,p$. Conveniently, \eqref{mini} is a standard weighted least-squares regression problem. Let $\boldsymbol{W}={\rm diag}\big\{K_h(x_1-x), \dots, K_h(x_n-x)\big\}$. Define
\begin{align}
\boldsymbol{Y}=\begin{bmatrix}
    y_1 \\
    \vdots \\
    y_{n}
\end{bmatrix},\>{\rm and}\>\boldsymbol{X}=\begin{bmatrix}
    1 & x_1-x & \dots & (x_1-x)^p\\
    \vdots & \vdots & \ddots & \vdots\\
    1 & x_n-x & \dots & (x_n-x)^p
\end{bmatrix}.\notag
\end{align}
Assuming invertibility of $\boldsymbol{X}'\boldsymbol{W}\boldsymbol{X}$, then $\hat{\boldsymbol{\beta}}=\left(\boldsymbol{X}^T\boldsymbol{W}\boldsymbol{X}\right)^{-1}\boldsymbol{X}^T\boldsymbol{W}\boldsymbol{Y}$. The estimate of the regression function at $x$ is $\hat{m}(x)=\hat{\beta}_0=\boldsymbol{e}^T_1\boldsymbol{\hat{\beta}}$, where $\boldsymbol{e}$ is the $(p+1)\times1$ vector with $1$ being the first entry and zero elsewhere. \cite{ruppert1994multivariate} studied the leading conditional bias and variance of the above estimator, but here we consider in detail only the local linear $(p=1)$ binned estimator. We make the following assumptions:
\begin{enumerate}
    \item[A1.] The function $m^{(2)}(\cdot)$ is continuous.
    \item[A2.] The kernel $K$ is symmetric and supported on $(-1,1)$. Also, $K$ has a bounded first derivative.
    \item[A3.] As $N\to\infty$ and $\omega\to0$, $Nh\to\infty$ and $\omega=o(h^2)$, where $N$ is sample size, $h$ is bandwidth and $\omega$ is bin width.
\end{enumerate}
It follows from the definition of the estimator that $E\{\hat{m}(x)\}\approx\boldsymbol{e}^T_1\left(\boldsymbol{X}^T\boldsymbol{W}\boldsymbol{X}\right)^{-1}\boldsymbol{X}^T\boldsymbol{W}\boldsymbol{M}$, where the vector $\boldsymbol{M}=\left\{m(x_1),\dots,m(x_n)\right\}^T$ contains the true regression function values at the each of the $x_i$'s. For local linear regression we have that
\begin{align}
\boldsymbol{X}=\begin{bmatrix}
    1 & x_1-x\\
    \vdots & \vdots \\
    1 & x_n-x
\end{bmatrix}.
\end{align}
Also, ${\rm Var}\{\hat{m}(x)\}=\boldsymbol{e}^T_1\left(\boldsymbol{X}^T\boldsymbol{W}\boldsymbol{X}\right)^{-1}\left(\boldsymbol{X}^T\boldsymbol{W}\boldsymbol{V}\boldsymbol{W}\boldsymbol{X}\right)\left(\boldsymbol{X}^T\boldsymbol{W}\boldsymbol{X}\right)^{-1}\boldsymbol{e}_1$, where $\boldsymbol{V} = {\rm Var}(\boldsymbol{Y})$ is a diagonal matrix with diagonal entries $m(x_i)/\omega N-m(x_i)^2/N$, $i=1,\dots,n$. Note that if $m$ is a linear function then the local linear estimator is exactly unbiased. To find the leading bias term for general function $m$, we review the result of local linear regression \cite{fan1993local} with the binning procedure discussed in \cite{hall1996accuracy}. This estimator uses only data on a single edge.

\begin{theorem}
\label{theorem1}
Suppose that $x$ is a point on the line segment of interest, and that A1, A2 and A3 hold. Let $\hat{m}(x)=\boldsymbol{e}^T_1\left(\boldsymbol{X}^T\boldsymbol{W}\boldsymbol{X}\right)^{-1}\boldsymbol{X}^T\boldsymbol{W}\boldsymbol{Y}$, then
\begin{align}
\label{b1}
    &E\left\{\hat{m}(x)-m(x)\right\}=\frac{1}{2}h^2\left\{\frac{\left(\sigma^2_K\right)^2-\left(\sigma^3_K\right)\sigma^1_K}{\sigma^2_K-\left(\sigma_K\right)^2}\right\}m^{(2)}(x)+O\left(\omega\right)+o\left(h^2\right)\\
    \label{v1}
    &{\rm Var}\left\{\hat{m}(x)\right\}=C(N,h,\omega,x)Q_K+o\left(\frac{1}{Nh}\right),
\end{align}
where $\sigma^i_K=\int u^i K(u)du$, $R^i_K=\int u^i K(u)^2 du$, 
\begin{align}
C(N,h,\omega,x) = \frac{1}{Nh}m(x)+\frac{\omega}{Nh}m(x)^2\>\>{\rm and}\>\>Q_K=\frac{R^0_K\left(\sigma^2_K\right)^2-2R^1_K\sigma^2_K\sigma^1_K+R^2_K\left(\sigma^1_K\right)^2}{\left\{\sigma^2_K-\left(\sigma^1_K\right)^2\right\}^2}.
\end{align}
\end{theorem}
The estimator has the following asymptotic distribution:
\begin{align}
\hat{m}(x)-m(x)-\frac{1}{2}h^2\left\{\frac{\left(\sigma^2_K\right)^2-\left(\sigma^3_K\right)\sigma^1_K}{\sigma^2_K-\left(\sigma_K^1\right)^2}\right\} m^{(2)}(x)\stackrel{d}{\to} N\left\{0,C(N,h,\omega,x)Q_K\right\}.
\end{align}
For an interior point $x$, we can simplify \eqref{b1} and \eqref{v1}. Note that in this case $\sigma^i_K=0$ for odd $i$, we have
\begin{align}
    &E\left\{\hat{m}(x)-m(x)\right\}=\frac{1}{2}h^2\sigma^2_Km^{(2)}(x)+O\left(\omega\right)+o\left(h^2\right)\\
    &{\rm Var}\left\{\hat{m}(x)\right\} =C(N,h,\omega,x)R^0_K+o\left(\frac{1}{Nh}\right).
\end{align}
The local linear estimator has better properties at the boundary than the kernel density estimator (KDE). Asymptotically, the local linear estimator's bias is of the same order, $h^2$, at a boundary point as at an interior point, where the bias would be of order $h$ for KDE. Even if $x$ is at the boundary of the density's support, since the local linear estimator fits a weighted least squares line through data near the boundary, if the true relationship is linear, this estimator will be exactly unbiased.\\
\\
Extension to higher order $p$ is straightforward, see \cite{ruppert1994multivariate}. For $p$ odd, the bias is of order $h^{p+1}$ everywhere.  When $p$ is even, the bias is of order $h^{p+2}$ away from the boundary but $h^{p+1}$ in the boundary region. By increasing the polynomial order from even to the next odd number, the order of the bias in the interior remains unchanged, but the bias simplifies. Kernel estimation corresponds to $p=0$. On the other hand, by increasing the polynomial order from odd to the next even number, the bias order increases in the interior. This effect is analogous to the bias reduction achieved by higher-order kernels.

\section{Test for joint equality at vertex}
\label{sec:test}
When multiple edges meet at a vertex,  there are as many estimates for $m(v)$. Consider a simple network consisting of three edges $e_1, e_2$ and $e_3$ and a vertex $v$ (see Figure~\ref{fig:simpleLinearNetwork}). One estimate is $\hat{m}_{e_1}(v)$, the local linear regression estimate 
of $m_{e_1}(v) := \lim_{x \to v,\ x \in e_1}m(x)$ 
using only data on $e_1$. The estimates $\hat{m}_{e_2}(v)$ and $\hat{m}_{e_3}(v)$ are constructed similarly. In this way, for a vertex connecting $J$ edges, we obtain $J$ estimates at that vertex. In this section, we construct a test to study whether the $m_{e_j}(v)$ for $j=1,\dots,J$ (or any subset) are equal.\\
\\
Let $v$ be a vertex and let $e_l$, $l=1,\dots,J$, be the edges connected by $v$. We want to test
\begin{align}
&H_0:\>m_{e_1}(x)=m_{e_2}(x)=\dots=m_{e_J}(x)\\
&H_1:\> {\rm Not}\> H_0.
\end{align}

\begin{theorem}
Let $\hat{\boldsymbol{m}}=\{\hat{m}_{e_1}(v),\dots, \hat{m}_{e_J}(v)\}^T$, and $\boldsymbol{C}$ be a contrast matrix such that $\boldsymbol{C1}=\boldsymbol{0}$. Under the null hypothesis, we have
\begin{align}
\label{test}
(\boldsymbol{C}\hat{\boldsymbol{m}})^T\boldsymbol{\Sigma}^{-1}\boldsymbol{C}\hat{\boldsymbol{m}}\stackrel{a}{\sim}\chi^2_{J-1},
\end{align}
where $\boldsymbol{\Sigma}=\boldsymbol{C}\, diag(V_{e_1},\dots,V_{e_J})\, \boldsymbol{C}^T$, and $V_{e_l}$ is the asymptotic variance of the $i$th estimator.  Here $\stackrel{a}{\sim}$
denotes ``asymptotically distributed as.''\label{theorem2}
\end{theorem}
We note that the test statistic \eqref{test} is invariant under the choice of contrast matrices. This is easy to see by noticing that the rows of $\boldsymbol{C}$ are linearly independent.

\section{Estimation with equal intercepts at the vertex}
\label{sec:estEqual}
If $H_0$ is not rejected, then evaluation points within the $h$-neighborhood of the vertex can be estimated using data from all neighboring edges, subject to $m_{e_1}(v)=\dots=m_{e_J}(v)$. The resulting estimator has a lower variance compared to estimating separately on each edge.

For now, we make the following assumptions:
\begin{enumerate}
    \item[B1.] No loop shorter than $2h$ in the network.
    \item[B2.] All edges are longer than $h$.
\end{enumerate}
If the $h$ neighborhood of the evaluation point $x$ is completely covered by an edge, then the estimator is the same as if we only use data from that edge, and its asymptotic bias and variance are given in Theorem 1. Next, we derive the estimator when the $h$ neighborhood $x$ contains a vertex. Note that, by assumption B1, there is a unique path between the evaluation point and any data point that is within its $h$-neighborhood, and by assumption B2, we need only consider the case where the neighborhood of $x$ contains exactly one vertex. For a fixed network, the above assumptions eventually will hold as $N\to\infty$ and $h\to0$. We will consider the effect of these assumptions on the implementation of the proposed estimator in Section~\ref{sec:practical}.\\
\\
We note that our estimator will have an additional bias when the test in Section 4 makes a type II error, that is, when the density is not continuous at the vertex but the test accepts that it is continuous. We will investigate this problem in Section~\ref{subsec:typeII}. 

\subsection{Deriving the estimator}
\begin{figure}[h]
\centering
\includegraphics[scale=0.7]{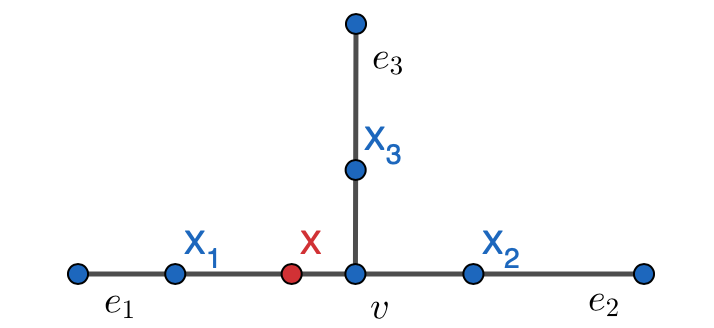}
\caption{Simple network.\label{fig:simpleLinearNetwork}}
\end{figure}

Let us first consider the simple network in Figure~\ref{fig:simpleLinearNetwork}, and consider a point $x\in e_1$ in the $h$-neighborhood of the vertex $v$, and data points $(x_1,y_1)$ on $e_1$, $(x_2,y_2)$ on $e_2$ and $(x_3,y_3)$ on $e_3$. Directional distance is used here. The evaluation point $x$ is the origin when measuring distance. Note that we will use directional distance in our proposed model. When we consider a path from a point on the network to another, we use the evaluation point $x$ as the origin. For example the distance between $x_1$ and $x$ is $x_1-x$ and the distance between $x_1$ and $v$ is $x_1-v$, which are both negative.
\\
Using a Taylor expansion up to first order, we can approximate the regression function on $e_1$ at $x_1$ as $m_{e_1}(x_1)\approx m_{e_1}(x)+m^{(1)}_{e_1}(x)(x_1-x)$ provided that all the required derivatives exist. For the regression function on $e_2$ at $x_2$, we have
\begin{align}
    m_{e_2}(x_2)&\approx
    m_{e_2}(v)+m^{(1)}_{e_2}(x)(x_2-v)\notag\\
    \label{s1}
    &=m_{e_1}(v)+m^{(1)}_{e_2}(x)(x_2-v)\\
    \label{s2}
    &\approx m_{e_1}(x)+m^{(1)}_{e_1}(x)(v-x) + m^{(1)}_{e_2}(v)(x_2-v),
\end{align}
where \eqref{s1} holds because that the regression functions from all edges are assumed to be equal at the vertex, \eqref{s2} is a Taylor expansion of $m_{e_1}(v)$ at $x$ on $e_1$, and $m^{(1)}_{e_j}(v)$ is defined as $\lim_{x\to v, x\in e_j} m_{e_j}^{(1)}(x)$. Similarly we have
\begin{align}
    m_{e_3}(x_3)\approx m_{e_1}(x)+m^{(1)}_{e_1}(x) (v-x) +m^{(1)}_{e_3}(v)(x_3-v).\notag
\end{align}
For a simple network with one vertex $v$ of degree $J$ and an evaluation point $x$ on $e_l$, we have for $x_i\in e_j$, $j=1,\dots,J$ and $j\ne l$, $m_{e_j}(x_i)\approx m_{e_l}(x)+m^{(1)}_{e_l}(x)(v-x) +m^{(1)}_{e_j}(v)(x_i-v)$, and for $x_i\in e_l$, $m_{e_l}(x_i)\approx m_{e_l}(x)+m^{(1)}_{e_l}(x)(x_i-x)$.\\
\\
Under the above construction, the regression functions agree at the vertex but can have different slopes there. Letting $\beta_0=m_{e_l}(x)$, $\beta_1(e_l)=m^{(1)}_{e_l}(x)$, and $\beta_1(e_j)=m^{(1)}_{e_j}(v)$ for $j\ne l$, these parameters are estimated by minimizing:
\begin{align}
\sum_{x_{il}\in e_l}\big\{y_{il}-\beta_0+\beta_1(e_l)(x_{il}-x)\big\}^2K_{il} +\sum_{j\ne l}\sum_{x_{ij}\in e_j}\big\{y_{ij}-\beta_0-\beta_1(e_l)(v-x)+\beta_1(e_j)(x_{ij}-v)\big\}^2K_{ij},\notag
\end{align}
where $K_{ij}=K\left(d_{ij}/h\right)$, and $d_{ij}$ is the network distance, i.e., the arc length, between $x_{ij}$ and $x$. In matrix notation, ~we~have
\begin{align}
\label{minimise}
    \left(\boldsymbol{Y}-\boldsymbol{X}\boldsymbol{\beta}\right)^T\boldsymbol{W}\left(\boldsymbol{Y}-\boldsymbol{X}\boldsymbol{\beta}\right),
\end{align}
where $\boldsymbol{\beta}=[\beta_0,\beta_1(e_1),\dots,\beta_1(e_J)]^T$ and $\boldsymbol{X}$ is a $n\times(J+1)$ matrix. The first column of $\boldsymbol{X}$ is identically $1$.  In the $j$th column, for $j>1$ and $j\ne l$, the $1+\sum_{i=1}^{j-1}n_i$ to $\sum_{i=1}^jn_i$ entries are $\{x_{ij}-v\}_{i=1,\dots,n_j}$, where $n_i$ is the number of bin centers on edge $i$, and the remaining entries are zero. In the $l$th column of $\boldsymbol{X}$, the $1+\sum_{i=1}^{l-1}n_i$ to $\sum_{i=1}^ln_i$ entries are $\{x_{il}-v\}_{i=1,\dots,n_l}$, and the remaining entries are $v-x$. We also note that $\boldsymbol{Y}=\{y_{ij}\}_{i=1,\dots,n_j; j=1\dots,J}$, and $\boldsymbol{W}={\rm diag}\{K_{ij}\}_{i=1,\dots,n_j; j=1\dots,J}$.\\
\\
The solution to the minimization problem \eqref{minimise} is $\hat{\boldsymbol{\beta}}=\left(\boldsymbol{X}^T\boldsymbol{W}\boldsymbol{X}\right)^{-1}\boldsymbol{X}^T\boldsymbol{W}\boldsymbol{Y}$,
and the estimate for the regression function at $x$ under the constraint $m_{e_1}(v)=\dots=m_{e_J}(v)$ is given~by
\begin{align}
\label{ehat}
\hat{m}(x)=\boldsymbol{e}_1^T\left(\boldsymbol{X}^T\boldsymbol{W}\boldsymbol{X}\right)^{-1}\boldsymbol{X}^T\boldsymbol{W}\boldsymbol{Y}.
\end{align}
Note that we can extend the above derivation to quadratic and higher order approximation, allowing different $m^{(p)}_{e_j}(v)$ for all $j$ under the constraint that $m^{p'}_{e_j}$ are equal for $p'<p$.

\subsection{Asymptotic properties -- the local linear case ($p=1$)}
In this subsection, we make the following assumptions:
\begin{enumerate}
    \item[C1.] The functions $m_{e_j}(v)$ are equal for all $j$.
    \item[C2.] The function $m^{(2)}_{e_j}(\cdot)$ for all $j$ are continuous on $e_j$, excluding the vertex.
    \item[C3.] The kernel $K$ is symmetric and supported on $(-1,1)$. Also, $K$ has a bounded first derivative.
    \item[C4.] The vertex $v$ is always in the $h$-neighborhood of $x$ as $h$ approaches 0, so $x$ must approach $v$ at least as fast as $h$ approaches 0.
    \item[C5.] As $N\to\infty$ and $\omega\to0$, $Nh\to\infty$ and $\omega=o(h^2)$, where $N$ is sample size, $h$ is bandwidth, and $\omega$ is bin width.
\end{enumerate}
Note that if C4 does not hold, then we are in the case already studied by \cite{ruppert1994multivariate}.

\begin{theorem}
Suppose that $x\in e_l$ is within the $h$-neighborhood of a vertex that connects $J$ segments, and that C1, C2, C3, C4 and C5 hold. Let $\hat{m}(x)=\boldsymbol{e}^T_1\left(\boldsymbol{X}^T\boldsymbol{W}\boldsymbol{X}\right)^{-1}\boldsymbol{X}^T\boldsymbol{W}\boldsymbol{Y}$, where the matrices are defined as above, then
\begin{align}
        &E\left\{\hat{m}(x)-m(x)\right\}=\frac{1}{2}\boldsymbol{e}^T_1\boldsymbol{U}_0^{-1}\left\{h^2\boldsymbol{R}_0+h(v-x)\left(\boldsymbol{R}_1-\boldsymbol{U}_1\boldsymbol{U}_0^{-1}\boldsymbol{R}_0\right)\right\}+o\left(h(v-x)\right)+o\left(h^2\right)+O(\omega),\notag\\
    &{and}\notag\\
    &{\rm Var}(\hat{m}(x))=\frac{1}{Nh}C(\omega,x)\boldsymbol{e}^T_1\boldsymbol{U}^{-1}_0\left[\boldsymbol{M}_0+\left(\frac{v-x}{h}\right)\left(\boldsymbol{M}_0\boldsymbol{U}^{-1}_0\boldsymbol{U}_1+\boldsymbol{M}_1-\boldsymbol{U}_1\boldsymbol{U}^{-1}_0\boldsymbol{M}_0\right)\right]\boldsymbol{U}^{-1}_0\boldsymbol{e}_1\notag\\
    &+o\left(\frac{1}{Nh}\right)+o\left(\frac{\omega}{Nh}\right)+o\left(\frac{v-x}{Nh^2}\right)+o\left(\frac{\omega(v-x)}{Nh^2}\right),\notag
\end{align}
where the matrices are defined as follows:
\begin{enumerate}
    \item $C(\omega,x)=m(x)-\omega m(x)^2$.
    \item $\boldsymbol{U}_0$ is symmetric, and its first row is $\sum_j\mu^{(j)}_0,\mu^{(1)}_1,\dots,\mu^{(J)}_1$ and its diagonal is $\sum_j\mu^{(j)}_0,\mu^{(1)}_2,$ $\dots,\mu^{(J)}_2$, where $\mu^{(j)}_i=\int_{e_j}u^iK(u)du$. All other entries are zero.
    \item $\boldsymbol{U}_1$ is symmetric, and its first row is $0,-\mu^{(1)}_0,\cdots,-\mu^{(l-1)}_0,$ $\sum_{j\ne l}\mu^{(j)}_0,-\mu^{(l+1)}_0,\cdots,-\mu^{(J)}_0$. Its $(l+1)$th row is $\sum_{j\ne l}\mu^{(j)}_0,\mu^{(1)}_1,\cdots,\mu^{(l-1)}_1,0,\mu^{(l+1)}_1,\cdots,\mu^{(J)}_1$ and its diagonal is $0,-2\mu^{(1)}_1,$ $\cdots,-2\mu^{(l-1)}_1,0,-2\mu^{(l+1)}_1,\cdots,-2\mu^{(J)}_1$. All other entries of $\boldsymbol{U}_1$ are zero.
    \item $\boldsymbol{R}_0$ is a column vector, its first entry is $\sum_{j\ne l}m^{(2)}_{e_j}(v)\mu^{(j)}_2+m^{(2)}_{e_l}(x)\mu^{(l)}_2$, and the rest are $m^{(2)}_{e_j}(v)\mu^{(j)}_3$, for $j=1,\cdots,J$.
    \item $\boldsymbol{R}_1$ is a column vector, its first entry is $-2\sum_{j\ne l}m^{(2)}_{e_j}(v)\mu^{(j)}_1$, and the rest, except the $(l+1)$th entry, are $-3m^{(2)}_{e_j}(v)\mu^{(j)}_2$, for $j\ne l+1$, and the $(l+1)$th entry is $\sum_{j\ne l}m^{(2)}_{e_{j}}(x)\mu^{(j)}_2$, for $j= l+1$.
    \item $\boldsymbol{M}_0$ is symmetric, its first row is $\sum_jR_0^{(j)}, R_1^{(1)},\cdots,R_1^{(J)}$, and its diagonal entries are $\sum_jR_0^{(j)}, R_2^{(1)},\cdots,R_2^{(J)}$. All the other entries of $\boldsymbol{M}_0$ are zero.
    \item $\boldsymbol{M}_1$ is symmetric, its first row is $0,-R^{(1)}_0,\cdots,-R^{(l-1)}_0,\sum_jR^{(j)}_0,-R^{(l+1)}_0,\cdots,-R^{(J)}_0$, its $(l+1)$th row is
    
    $\sum_jR^{(j)}_0,R^{(1)}_1,\cdots,R^{(l-1)}_1,0,$ $R^{(l+1)}_1,\cdots,R^{(J)}_1$, and its diagonal is $0,-R^{(1)}_1,\cdots,$ $-R^{(l-1)}_1,0,-R^{(l+1)}_1,\cdots,-R^{(J)}_1$. All other entries of $\boldsymbol{M}_1$ are zero.
\end{enumerate}
\end{theorem}
Note that for $v$ to be in the $h$-neighborhood of the evaluation point $x$, we need $|v-x|<h$, and for $v$ to stay in the $h$-neighborhood of the evaluation point asymptotically, we need $x$ to approach $v$ at least as fast as $h$ approaches $0$. So $(v-x)/h$ is small and approaches zero as $N\to\infty$.

\subsection{Estimation with equal first derivatives at the vertex}
We have studied the asymptotic properties of local piecewise linear regression estimator at evaluation points that are within the $h$-neighborhood of a vertex, under the assumption that ${m}_{e_j}(v)$, $j=1,\dots,J$, agree at the vertex. Similarly, we can construct an asymptotic test for the joint equality of the limits of ${m}^{(1)}_{e_j}(x)$, $j=1,\dots,J$, as $x\to v$ along multiple edges. If we accept the null hypothesis that the first derivatives are equal at the vertex, this constraint should be added to the estimation procedure. For example, in Figure~\ref{fig:simpleLinearNetwork}, consider an evaluation point $x\in e_1$ and a data point $x_2\in e_2$.  Using local linear approximations, we have
\begin{align}
    m_{e_2}(x_2)&\approx
    m_{e_2}(v)+m^{(1)}_{e_2}(v)(x_2-v)\notag\\
    &=m_{e_1}(v)+m^{(1)}_{e_2}(v)(x_2-v)\notag\\
    &\approx m_{e_1}(x)+m^{(1)}_{e_1}(x)(v-x) + m^{(1)}_{e_2}(v)(x_2-v)\notag\\
    \label{highorder}
    &\approx m_{e_1}(x)+m^{(1)}_{e_1}(x)(v-x) + m^{(1)}_{e_1}(v)(x_2-v).
\end{align}
The regression function at $x_3\in e_3$ has a similar expansion. In the last line we used $m^{(1)}_{e_2}(v)=m^{(1)}_{e_1}(v)$, and expand $m^{(1)}_{e_1}(v)$ around the evaluation point $x$ on $e_1$. However to estimate first derivatives, one should use at least local quadratic polynomials.

\section{Practical Issues}\label{sec:practical}
\subsection{Estimation with large bandwidth}
In this section, we will describe a more complex situation. When there are loops in the network, we use the shortest path between the evaluation point and a data point, and the distance between the points is defined by the shortest path distance. When there are multiple vertices in the $h$-neighborhood of an evaluation point, and vertices have different results from the joint equality tests, we will show that only data points that have {\it direct access} to the evaluation point contribute to the estimation. If $x$ is the evaluation point,  $x_i$ is a data point that is within the $h$-neighborhood of $x$, and $v_1,\dots,v_n$ are the vertices on the shortest path between $x$ and $x_i$, we say that $x_i$ has {\it direct access} to $x$ if the H$_0$ (Theorem 2) is accepted at all $v_i$, $i=1,\cdots,n$. We will demonstrate this via the following example.\\
\begin{figure}[h]
\label{fig:simpleLinearNetwork2}
\centering
\includegraphics[scale=0.65]{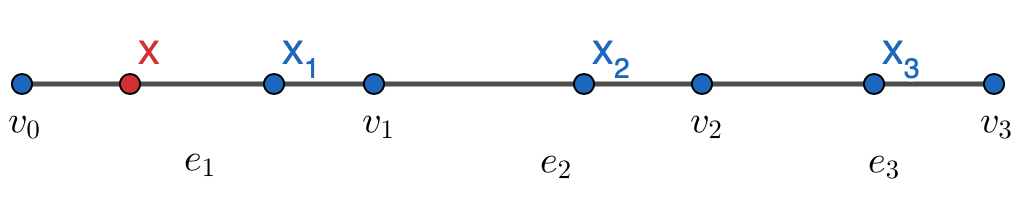}
\caption{Simple Linear Network.}
\end{figure}
\\
Here $x$ is the evaluation point, and $x_1$, $x_2$ and $x_3$ are data points on edges $e_1$, $e_2$ and $e_3$ respectively. We consider two scenarios:
\begin{enumerate}
    \item $H_0$ is accepted at $v_1$ and $v_2$,
    \item $H_0$ is accepted at $v_1$ and rejected at $v_2$.
\end{enumerate}
In the first scenario, by Taylor expansions, we have
\begin{align}
\label{one}
  &m_{e_1}(x_1)\approx m_{e_1}(x)+(x_1-x)m^{(1)}_{e_1}(x)\\
  \label{two}
&m_{e_2}(x_2)\approx m_{e_2}(v_1)+(x_2-v_1)m^{(1)}_{e_2}(v_1)  \\
\label{three}
  &m_{e_3}(x_3)\approx m_{e_3}(v_2)+(x_3-v_2)m^{(1)}_{e_3}(v_2).
\end{align}
Now, since $H_0$ is accepted at $v_1$ and $v_2$, we have $m_{e_1}(v_1)=m_{e_2}(v_1)$ and $m_{e_2}(v_2)=m_{e_3}(v_2)$, and \eqref{two} and \eqref{three} become
\begin{align}
&m_{e_2}(x_2)\approx m_{e_1}(x)+(v_1-x)m^{(1)}_{e_1}(x)+(x_2-v_1)m^{(1)}_{e_2}(v_1)  \notag\\
\label{four}
  &m_{e_3}(x_3)\approx m_{e_2}(v_1)+(v_2-v_1)m^{(1)}_{e_2}(v_1)+(x_3-v_2)m^{(1)}_{e_3}(v_2)\\
  &\approx m_{e_1}(x)+(v_1-x)m^{(1)}_{e_1}(x)+(v_2-v_1)m^{(1)}_{e_2}(v_1)+(x_3-v_2)m^{(1)}_{e_3}(v_2).\notag
\end{align}
The  problem of minimizing $\sum_{i,j}\{y_{ij}-m_{e_j}(x_i)\}^2K_{ij}$ can be set up in the same way as before. However, in the second scenario, since we do not have $m_{e_2}(v_2)=m_{e_3}(v_2)$, the problem becomes minimization of $\sum_{i,j=1,2}\{y_{ij}-m_{e_j}(x_i)\}^2K_{ij}+\sum_{i}\{y_{i3}-m_{e_3}(x_i)\}^2K_{i3}$.  We see that the estimation of $m_{e_1}(x)$ only depends on the first summation. In other words, only data points that have direct access to the evaluation point contribute to the estimation of $m(x)$.

\subsection{Bin width}
For fixed sample size, we will let the bin width $\omega\to0$. For simplicity, we show how this affects the calculation of $\left(\omega\boldsymbol{X}^T\boldsymbol{W}\boldsymbol{X}\right)^{-1}\omega\boldsymbol{X}^T\boldsymbol{W}\boldsymbol{Y}$. In the case where the regression function is estimated on each edge individually, when $\omega\to0$, to calculate $\omega\boldsymbol{X}^T\boldsymbol{W}\boldsymbol{X}$ for $x$ near the vertex, we use that the Riemann sum converges to the integral as $\omega\to0$:
\begin{align}
    \omega\sum^n_{i=1}(x_i-x)^pK_h(x_i-x)\to\int_{L} u^p K(u)du=\mu^p(c),\notag
\end{align}
where $x=ch$, $0\le c\le1$, and $\mu^p(c)=\int^1_{-c} u^p K(u)du$ is the $p$th truncated moment. Note that when the evaluation point is the vertex itself, $c=0$. The calculation of the term $\omega\boldsymbol{X}^T\boldsymbol{W}\boldsymbol{Y}$ requires 
\begin{align}
    \omega\sum_{i=1}^{n}y_{i}(x_{i}-x)^{p}K_h\left({x_{i}-x}\right) =  \sum_{i=1}^{n}\frac{c_{i}}{N}(x_{i}-x)^{p}K_h\left({x_{i}-x}\right)\to \frac{1}{N}\sum_{i=1}^N(z_i-x)^pK_h\left({z_i-x}\right).\notag
\end{align}
Here $y$ is bar height and $\omega$ is bar width, so $y\, \omega$ is bar area, which equals to $c/N$, where $c$ is bar count and $N$ is sample size. Let $z_i$ be the location of $i$th observation. As $\omega\to0$, eventually each bin is either empty or has exactly one observation in it. Therefore bin count $c_{i}$ is $0$ for most bins, and $1$ for bins that contain one observation, and the location of that observation becomes the bin center in the limit.\\
\\
When the joint equality constraint is added, to approximate $\omega\boldsymbol{X}^T\boldsymbol{W}\boldsymbol{X}$ for $x\in e_l$ near the vertex, we also need to calculate, for $j\ne l$, $x_{ij}\in e_j$,
\begin{align}
    &\omega\sum^n_{i=1}(x_{ij}-v)K_h(x_{ij}-x)\to h\mu^1(c)-(v-x)\mu^0(c)\notag\\
    &\omega\sum^n_{i=1}(x_{ij}-v)^2K_h(x_{ij}-x)\to h^2\mu^2(c)-2h\mu^1(c)+(v-x)^2\mu^0(c)\notag.
\end{align}
Binning with $\omega$ fixed might still be used with very large sample size.

\section{Implementation}\label{sec:implement}

\subsection{Simulation studies}
\label{subsec:simulations}
\begin{figure}
\centering
\includegraphics[scale=0.825]{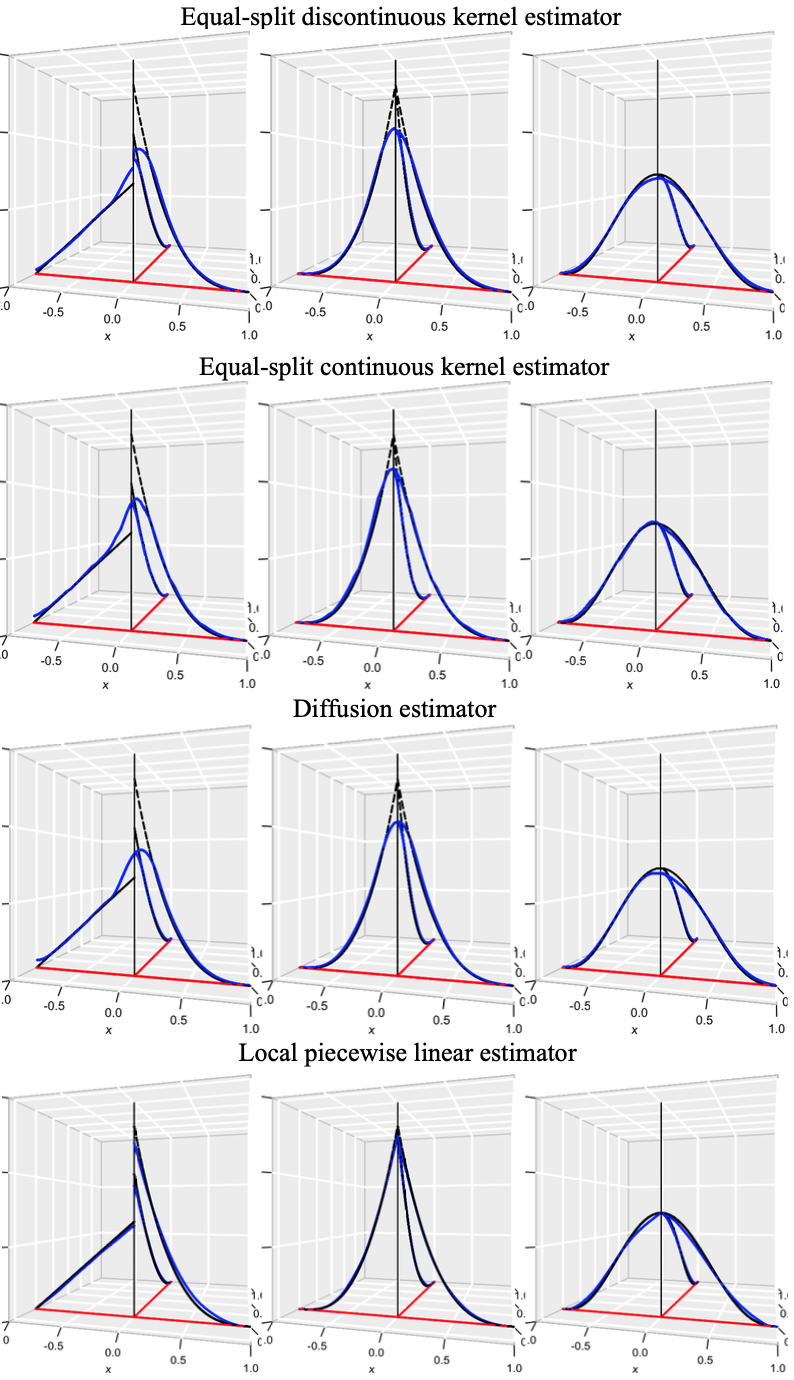}
\caption{Equal-split discontinuous and continuous kernel estimator and diffusion estimator applied to the three cases. From left to right: case I (discontinuous density), II (continuous density with discontinuous first derivative) and III (density and its first derivative are both continuous). The network is in red, and the true density functions over the network are the black dashed lines. The blue lines are the estimates. \label{fig:twelveEst}}
\end{figure}

\begin{figure}
\centering
\includegraphics[scale=0.6]{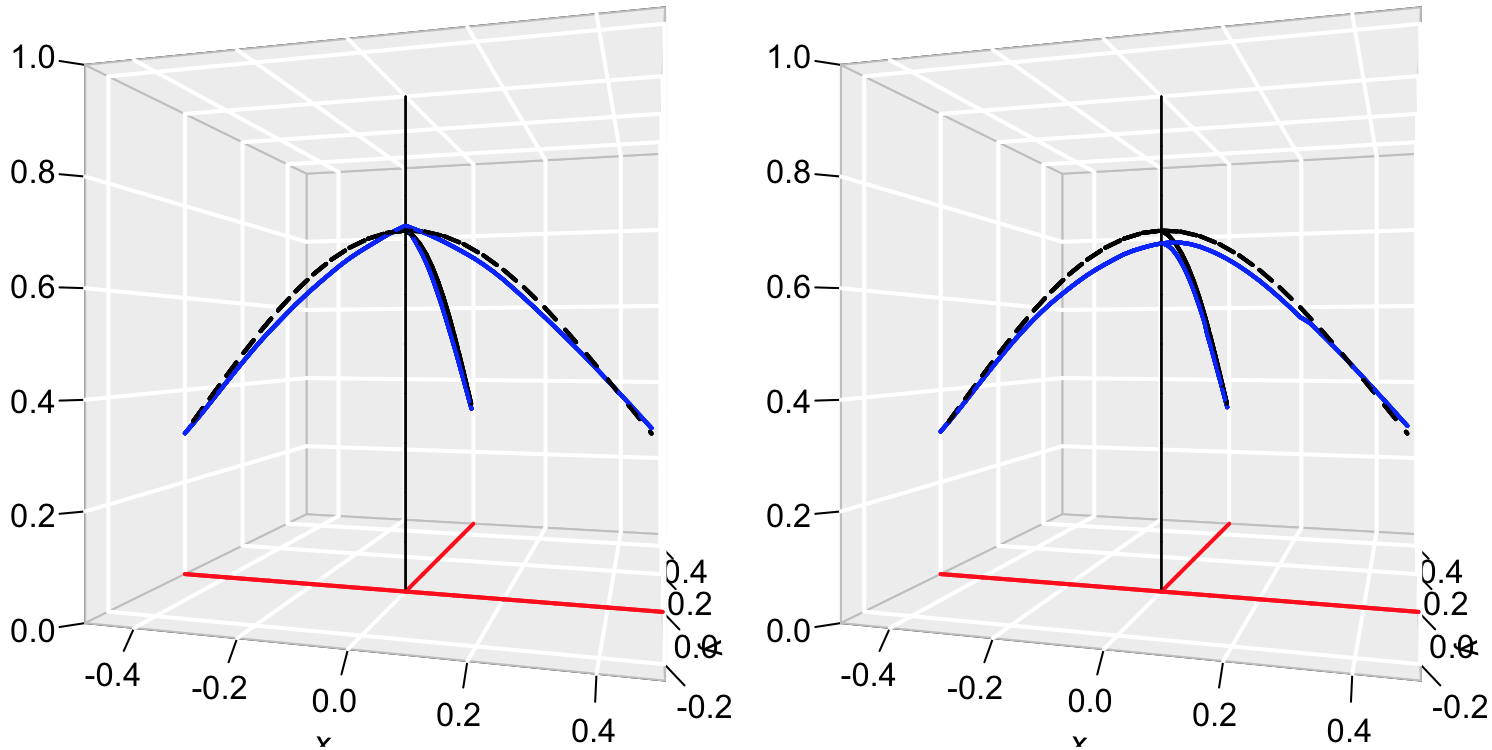}
\caption{The network is in red, and the true density functions over the network are the black dashed lines. {\bf Left:} local piecewise linear estimation subject to joint equality of the regression functions at the vertex. {\bf Right:} local piecewise linear estimation subject to that the first derivatives are jointly equal at the vertex.\label{fig:twoEstimates}}
\end{figure}

Let us consider the simple network shown in Figure~\ref{fig:simpleLinearNetwork}: three edges meet at a vertex. We propose three interesting cases on this network using the beta distribution. The three cases that we report here are
\begin{enumerate}
    \item We simulate $500$ points on each edge, with the vertex being the origin, from $Beta(1,2)$, $Beta(1,3)$, and $Beta(1,4)$ respectively.
    \item We simulate $500$ points on each edge, with the vertex being the origin, from $Beta(1,4)$.
    \item For each edge, with the vertex being the origin, we simulate $500$ points from the truncated (from $0.5$ to $1$) $Beta(4,4)$. Then the points are shifted to the vertex by $0.5$ and finally multiplied by $2$, so the support of the true density on each edge is from the vertex to a point that is unit distance away from the vertex. 
\end{enumerate}
The true density function over the network is normalized so it integrates to $1$. Our simulation study focuses on demonstrating that the proposed estimator can accommodate various behaviors of the true regression function (the true density function), especially at the vertex, namely, discontinuous (Case I), continuous with discontinuous first derivative (Case II), and continuous with continuous first~ derivative (Case III).\\
\\
Simulation results for one dataset are shown in Figure~\ref{fig:twelveEst}. The result for local piecewise linear density is presented in the last row. The network is in red, and the true density functions over the network are the black dashed lines. The blue lines are our estimates. The first column is case I, where the joint equality test for the regression functions at the vertex is rejected, and our estimate is equivalent to local linear regression on each edge. The second column is case II. We first fitted a local linear regression model on each edge, and tested joint equality at the vertex. Since we fail to reject the null hypothesis, we re-estimated the regression function in the $h$-neighborhood of $v$ using data from all neighboring edges subject to the equality constraint. Turning to case III in the third column, the estimate is also subject to the joint equality constraint.\\
\\
In case III, in addition to joint equality of the regression functions at the vertex, joint equality of their first derivatives is also established. For better viewing and comparison between different amounts of smoothness at the vertex, Figure~\ref{fig:twoEstimates} zooms in at the vertex. Estimation in left panel of Figure~\ref{fig:twoEstimates} is subject to only one constraint (joint equality of the regression functions at the vertex). However, after testing, by adding the constraint that the first derivatives are jointly equal, we see in the right panel of Figure~\ref{fig:twoEstimates} that the estimated curve is smoother over the vertex. This is desired because the true density function is continuous and has continuous first derivative over the vertex. \\
\\
We should note that any sufficiently small bin width will do, but selection of the bandwidth takes more care. We note that cross-validation could be used for bandwidth selection, and more computationally efficient bandwidth selector for polynomial regression on networks is under investigation.

\subsection{Comparison with existing methods} Let us consider again the three cases described in Section~\ref{subsec:simulations} and apply the equal-split discontinuous kernel estimator (ESDK) \cite{okabe2009kernel}, the equal-split continuous kernel estimator (ESCK) \cite{okabe2012spatial}, and the diffusion estimator (DE) \cite{mcswiggan2017kernel} to each case. \\
\\
ESDK \cite{okabe2009kernel} modifies \eqref{kde}, so it conserves mass. The algorithm makes a copy of the kernel function for each evaluation point, confined to the line edge containing that point. At each fork, the remaining tail of the kernel is split equally between the outgoing edges. Suppose there are $J-1$ outgoing edges, each outgoing edge receives a copy of the kernel tail weighted by $1/(J-1)$. ESCK \cite{okabe2012spatial} uses another modified version of \eqref{kde}. At each fork, with $J-1$ outgoing edges, each outgoing edge receives a copy of the kernel weighted by $2/J$, while the incoming segment receives a copy with the negative weight $2/J-1$. DE \cite{mcswiggan2017kernel} uses the estimator $\hat{f}(u)=\sum^N_{i=1}K_t(u|x_i)/N$,
where $K_t(u|z)$ is the heat kernel on a network \cite{botev2010kernel}. Intuitively, $K_t(u|z)du$ is the probability that a Brownian motion on the network, started at location $z\in L_T$ at time $0$, will fall in the infinitesimal interval of length $du$ around the point $u$ at time $t$. The bandwidth parameters for ESDK, ESCK, and DE are selected by cross validation. \\
\\
Figure~\ref{fig:twelveEst} illustrates the performances of ESDK, ESCK and DE, compared to LPLR. Columns one, two and three correspond to cases I, II and III, respectively. 
For case I (discontinuous density), both ESCK and DE produce a continuous estimate at the vertex. ESDK, however, produces a discontinuous estimate, but the bias is still significant. Although ESDK is not a standard kernel density estimator, one should note that the asymptotic bias of kernel density estimation has order of $h$ on the boundary, whereas local linear estimation has order of $h^2$. The significant bias at the vertex can be also due to the fact that ESDK uses data from all neighboring edges as the evaluation point approaches the vertex, so it overestimates low density edges and underestimate high density edges. \\
\\
For case II (continuous density with discontinuous first derivative), ESDK produces a discontinuous estimate. Although ESCK and DE produce continuous estimates, their bias are considerably higher compared to local piecewise polynomial estimation. This is due to lower order of the asymptotic bias of the kernel estimator near a vertex. \\
\\
For case III (the density and its first derivative are both continuous) the vertex behaves like an interior point.  While ESDK still produces a discontinuous estimate, ESCK and DE are comparable to local piecewise linear estimation, because the asymptotic bias of kernel density estimation and local linear estimation are both of order $h^2$ at an interior point. However, ESCK and DE do not estimate the derivatives of the estimated curves as the evaluation point approaching a vertex from multiple directions. Hence, these methods are inadequate when the derivatives are of interest.\\
\\
Finally we report the bias, standard deviation, and mean squared error of the proposed local (piecewise) polynomial regression compared to all existing methods. In each case (I, II and III), the result is based on simulation of $100$ data sets, and each data set consists of $1000$ data points on each edge. Bias, standard deviation and mean squared error are reported at the vertex as it is approached from $e_2$ (see Figure 2). The simulation result is summarised in Table 1. We see that ESDK, ESCK and DE only produce comparable results when the vertex behaves like an interior point, whereas LPLR is superior in all other cases. Simulations are run on a Macbook Pro Mid 2015 with 2.5 GHz Intel Core i7. Average time (Case I, II and III) for LPLR, ESDK, ESCK and DE are 6.0082 secs, 4.3217 hrs, 6.4872 hrs and 6.6266 secs for all $1000$ datasets. ESDK, ESCK and DE are implemented in the \texttt{R} package \texttt{spatstat}. We note that the proposed local linear estimator requires at least moderately large data sets in order to produce good models. But for a fixed network and a fixed number of evaluation points, since the optimal bandwidth is of order $n^{-1/5}$, the computation time only scales sub-linearly with sample size. More generally, the computation time is $O(n^{4/5})$ per evaluation point. If the number of evaluation points is $O(n^{1/5})$, then the computation time grows like $n$.
\begin{table}[]
\centering
\begin{tabular}{@{}llllllllll@{}}
\toprule
 & \multicolumn{3}{c}{Case I} & \multicolumn{3}{c}{Case II} & \multicolumn{3}{c}{Case III} \\ \midrule
 \hline
 & Bias & SD & MSE & Bias & SD & MSE & Bias & SD & MSE \\ \midrule
 \hline
LPLR & $-$0.0621 & 0.0593 & 0.0074 & $-$0.0629 & 0.0423 & 0.0057 & $-$0.0971 & 0.0123 & 0.0096 \\
ESDK & $-$0.4481 & 0.0196 & 0.2011 & $-$0.2711 & 0.0229 & 0.0741 & $-$0.1062 & 0.0121 & 0.0114 \\
ESCK & $-$0.4956 & 0.0184 & 0.2459 & $-$0.2312 & 0.0301 & 0.0542 & $-$0.1324 & 0.0100 & 0.0176 \\
DE & $-$0.4670 & 0.0172 & 0.2180 & $-$0.2438 & 0.0173 & 0.0597 & $-$0.1140 & 0.0116 & 0.0131 \\ \bottomrule
\hline
\end{tabular}
\caption{Bias, standard deviation, and mean squared error of the proposed local (piecewise) polynomial regression compared to all existing methods. The result is based on $100$ simulations of Cases I, III and III with $1000$ data points on each edge. Bias, standard deviation and mean squared error are reported at the vertex as it is approached from $e_2$ in Figure~\ref{fig:simpleLinearNetwork}.}
\label{t1}
\end{table}

\subsection{Additional bias}\label{subsec:typeII}
The proposed method is subject to additional error because of the limitations of the test described in Section \ref{sec:test}. For a large network, we would have to consider a large set of statistical tests simultaneously, and multiple testing problem would likely occur. It means that among the vertices at which the true density is continuous, the test rejected continuity at least one of them. However the main focus of this paper is the estimation of density function at individual locations, and the bias introduced by mistaking a continuous vertex as a discontinuous vertex is just the boundary bias of a local polynomial regression estimator. If one wants to count the number of  vertices at which the density function is discontinuous, numerous correction techniques have been proposed in the literature, such as the Holm-Bonferroni method \cite{holm1979simple}. Hence we will focus on type II error.\\
\\
Our estimator will also have an additional bias when the test that the density is continuous at a vertex makes a type II error, that is, when the density is not continuous but the test accepts that it is continuous. The size of additional bias introduced by a type II error is less clear. We investigate this problem through simulations. For simplicity, let the origin be a vertex, and let the two unit-length edges meet at the vertex. On the left edge, we first sample 1000 points from Beta$(1,\beta_l)$, then multiply the sample by $-1$, so the data points on the left edge range from $-1$ to $0$. On the right edge, we draw 1000 point from Beta$(1,\beta_r)$. Pairs of $(\beta_l, \beta_r)$ are given in the first column of Table~\ref{my-label}. Type II error rate is the probability that the test (Theorem~\ref{theorem2}) accepts the null hypothesis that the true density is continuous at a vertex when it is not. We perform $3000$ simulations of $1000$ data points per edge for each pair of $(\beta_l, \beta_r)$. We report the bias, standard deviation and mean squared error at the vertex approached from the right edge. The result is summarized in Table~\ref{my-label}. We see that when type II error rates are large, LPLR under the continuity constraint produces little additional bias due to small gaps between $\beta_l$ and $\beta_r$. Similarly, when $\beta_l$ and $\beta_r$ are far apart, there is also little additional bias due to small type II error rates. On the other hand, when type II error rate is in the mid range, for example from $0.6$ to $0.7$, there is considerable amount of additional bias at the vertex. Hence in the bias column, we should see bigger (in magnitude) biases in the middle rows, and as we approach the top and bottom rows, biases become smaller (in magnitude). Finally we note that the top row has bigger bias than the bottom row, despite that the type II error rate is practically zero. This is due to that local linear regression generally has bigger bias for functions with greater curvature (Beta$(1,4.5)$ compared to Beta$(1,4)$).
\begin{table}[]
\centering
\begin{tabular}{@{}lllll@{}}
\toprule
$(\beta_l, \beta_r)$ & Type II error rate & Bias & SD & MSE \\ \midrule
\hline
(3.5, 4.5) & 0.0056 & $-$0.0796 & 0.1480 & 0.0209 \\
(3.55, 4.45) & 0.0227 & $-$0.0786 & 0.1482 & 0.0199 \\
(3.6, 4.4) & 0.0540 & $-$0.0811 & 0.1458 & 0.0215 \\
(3.65, 4.35) & 0.1283 & $-$0.0833 & 0.1453 & 0.0207 \\
(3.7, 4.3) & 0.2453 & $-$0.0861 & 0.1487 & 0.0210 \\
(3.75, 4.25) & 0.3906 & $-$0.0925 & 0.1454 & 0.0203 \\
(3.8, 4.2) & 0.6030 & $-$0.1017 & 0.1398 & 0.0183 \\
(3.85, 4.15) & 0.7633 & $-$0.1040 & 0.1245 & 0.0158 \\
(3.9, 4.1) & 0.8863 & $-$0.0956 & 0.1135 & 0.0136 \\
(3.95, 4.05) & 0.9460 & $-$0.0781 & 0.1038 & 0.0127 \\
(4, 4) & NA & $-$0.0631 & 0.1013 & 0.0140 \\ \bottomrule
\hline
\end{tabular}
\caption{Type II error, bias, standard deviation and mean squared error at the vertex approached from the right edge, based on $3000$ simulations of $1000$ data points per edge in each scenario.}
\label{my-label}
\end{table}
\subsection{Application to real data} In this section we apply the proposed local piecewise polynomial regression to dendrite data. The data was collected by the Kosik Lab, UC Santa Barbara, and first analyzed by \cite{baddeley2014multitype} and \cite{jammalamadaka2013statistical}. Dendrites are branching filaments which extend from the main body of a neuron (nerve cell) to propagate electrochemical signals. Spines are small protrusions on the dendrites. The network shown in Figure~\ref{fig:dendriticSplines} is one of the ten dendritic trees of this neuron. A dendritic tree consists of all dendrites issuing from a single root branching off the cell body; each neuron typically has 4 to 10 dendritic trees. This example was chosen because it is large enough to demonstrate our techniques clearly, without being too large for graphical purposes. The events on the network are the locations of 566 spines observed on one branch of the dendritic tree of a rat neuron. We will show the result of applying the existing methods and the proposed local (piecewise) polynomial estimator to the dendrite data of Figure~\ref{fig:dendriticSplines}. We show density estimates at vertices A, B, C and D. We used a fixed bandwidth of $9$ microns (the network has a total length of $1933.653$ microns), and we will leave bandwidth selection and variable bandwidths to future projects. Unlike ESDK, ESCK and DE, the continuity of the LPLR estimates at a vertex is only imposed when there is evidence that the density is continuous there. Since the proposed method is particularly advantageous near vertices, we will mostly focus on that region.\\
\\
\begin{figure}[h]
\centering
\includegraphics[scale=0.7]{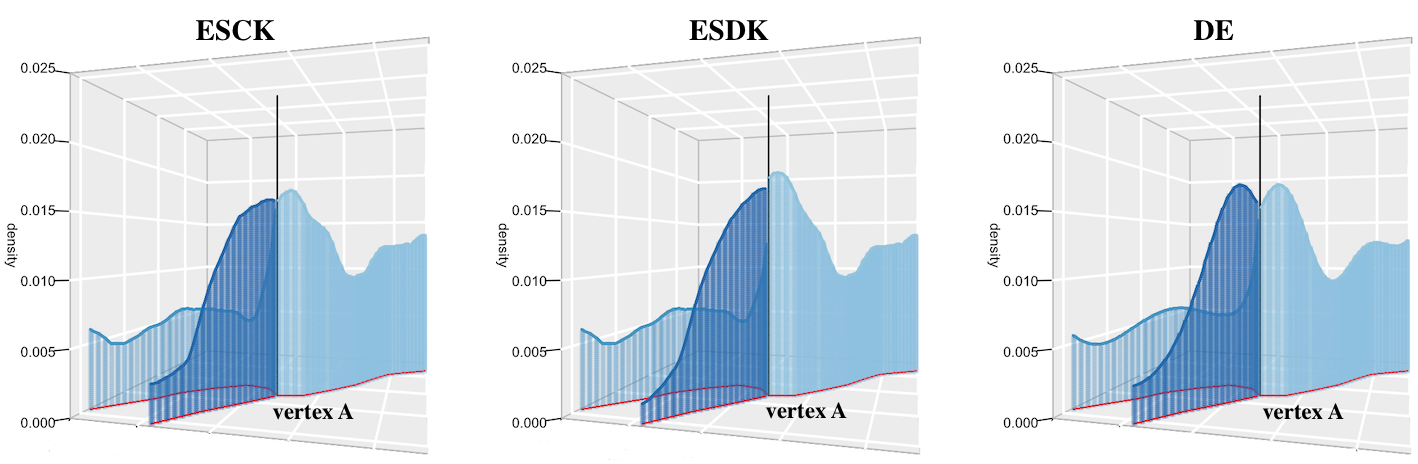}
\caption{Density estimation near vertex A, by ESCK, ESDK and DE. The network is in red and the estimates are in blue.}\label{est_a_exist}
\end{figure}
\begin{figure}[h]
\centering
\includegraphics[scale=0.7]{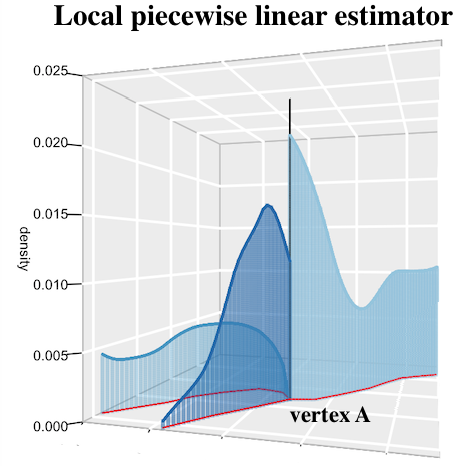}
\caption{Density estimation near vertex A, by local piecewise estimator. The network is in red and the estimates are in blue.}\label{est_a_new}
\end{figure}
Figure \ref{est_a_exist} shows density estimates near vertex A, using ESCK, ESDK and DE. We used a fixed bandwidth of $9$ microns, and an Epanechnikov kernel in ESCK and SEDK. Both ESCK and DE produced continuous estimates at the vertex. Although ESCK produced discontinuity at the vertex, the estimation clearly used data from all edges. Consequently the estimated density on the background edge near the vertex is overly inflated. In Figure \ref{fig:dendriticSplines}, the data suggests that the density function has different behaviours when approaching vertex A from different directions. However none of the exisiting methods can characterise this. Next we apply the proposed piecewise local linear regression approach, and the result is in Figure \ref{est_a_new}. Since the data does not support continuity at vertex A, no further constraint estimation is required.\\
\begin{figure}[h]
\centering
\includegraphics[scale=0.7]{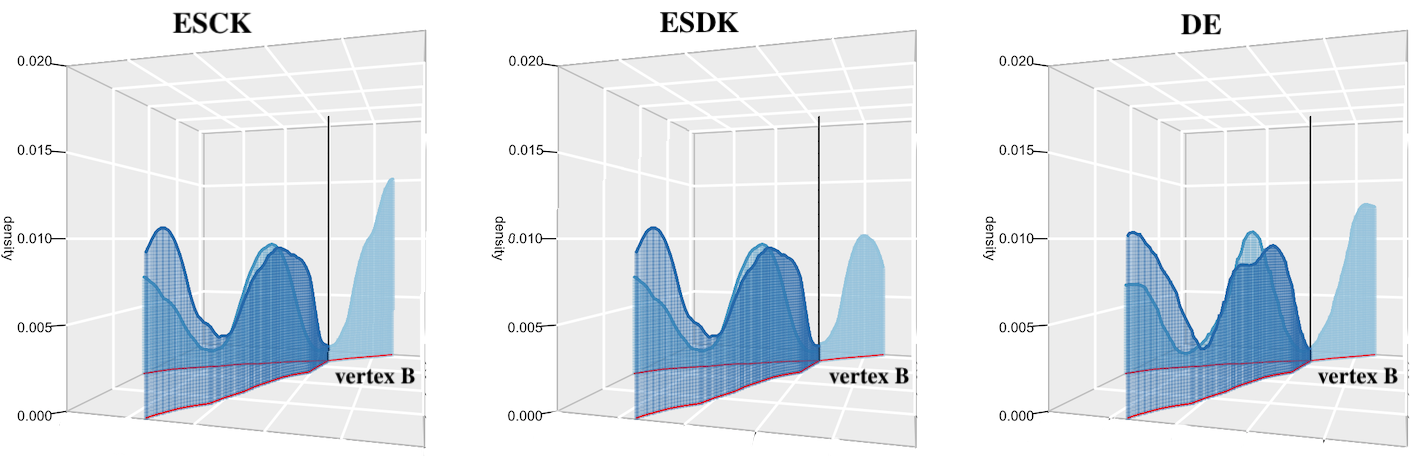}
\caption{Density estimation near vertex B, by ESCK, ESDK and DE. The network is in red and the estimates are in blue.}\label{est_b_exist}
\end{figure}
\begin{figure}[h]
\centering
\includegraphics[scale=0.75]{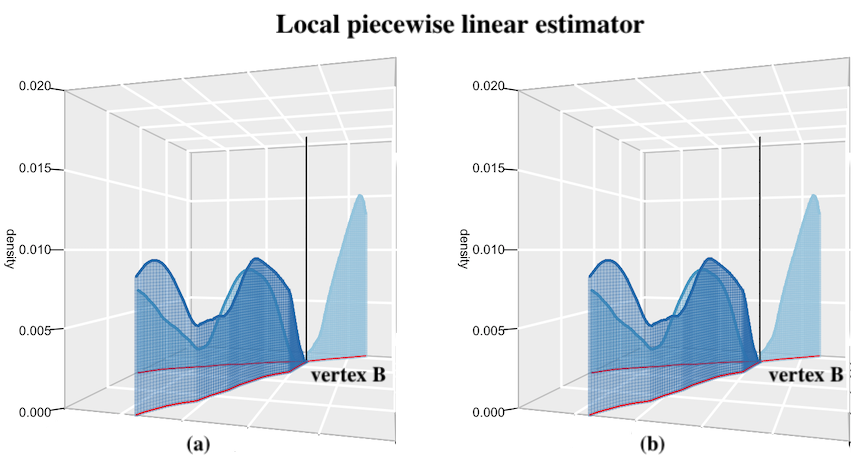}
\caption{(a): Density estimation near vertex B by local piecewise estimator using only the data on each edge. (b):  Density estimation near vertex B by local piecewise estimator using data from all three edges subject to the constraint that the densities on the left two edges are equal at the vertex. The network is in red and the estimates are in blue.}\label{est_b_new}
\end{figure}
\\
Next we look at vertex B in Figure \ref{fig:dendriticSplines}. Unlike vertex A, spines around vertex B seem to become more sparse as approaching the vertex from all directions, especially from the two edges on the left. Apply ESCK, ESDK and DE, we get the estimates in Figure \ref{est_b_exist}. ESCK and DE produced continuous estimates at vertex B. Under these two approaches, the estimated density is continuous at the vertex over any pair of edges, i.e. $\hat{f}(x)$ is continuous as $x$ travels from $e_i$ to $e_j$, where $e_i$ and $e_j$ are two edges connected by the vertex. However continuity may only exist over a subset of edges, and ESCK, ESDK and DE are unable to provide such flexibility. By the proposed piecewise local linear estimation, we first estimate the density on each edge individually using only the data on that edge. This is panel (a) in Figure \ref{est_b_new}. By the testing procedure discussed in Section \ref{sec:test}, there's strong evidence in the data that the densities from the left two edges are equal when approaching the vertex. Following the constrained estimation in Section \ref{sec:estEqual},  we re-estimate the densities on the two edges by local piecewise estimator using data from all three edges subject to the constraint that the densities on the left two edges are equal at the vertex. The result is in panel (b). Suppose $e_j$ for $j=1,\cdots,J$ are connected by vertex $v$, let $f_{e_j}(x)$ be the density over $e_j$. Using the proposed piecewise local linear estimation, we can test for the equality of $f_{e_j}(x)$ for any subset of $j\in\{1,\cdots,J\}$,as $x\to v$, and achieve the desired continuity as suggested by the data. On the other hand, under ESCK, ESDK and DE, the continuity at a vertex is decided by the user prior to model fitting, and the decision making is not data-driven.\\
\\
Similarly, for vertex C,  under piecewise local linear estimation, there's evidence in the data that densities from the two edges in the foreground are equal when approaching the vertex. Figure \ref{est_c_old} shows the estimated densities near vertex C, by ESCK, ESDK and DE, while piecewise local linear estimation is in Figure \ref{est_c_new}.
\begin{figure}[h]
\centering
\includegraphics[scale=0.7]{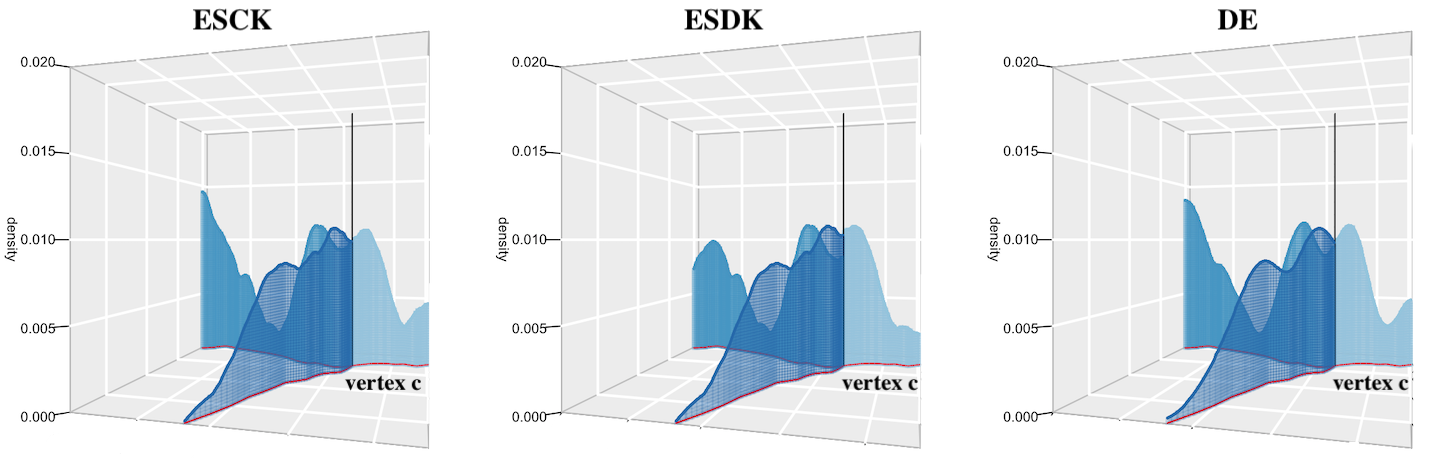}
\caption{Density estimation near vertex C, by ESCK, ESDK and DE. The network is in red and the estimates are in blue. The network is in red and the estimates are in blue.}\label{est_c_old}
\centering
\includegraphics[scale=0.7]{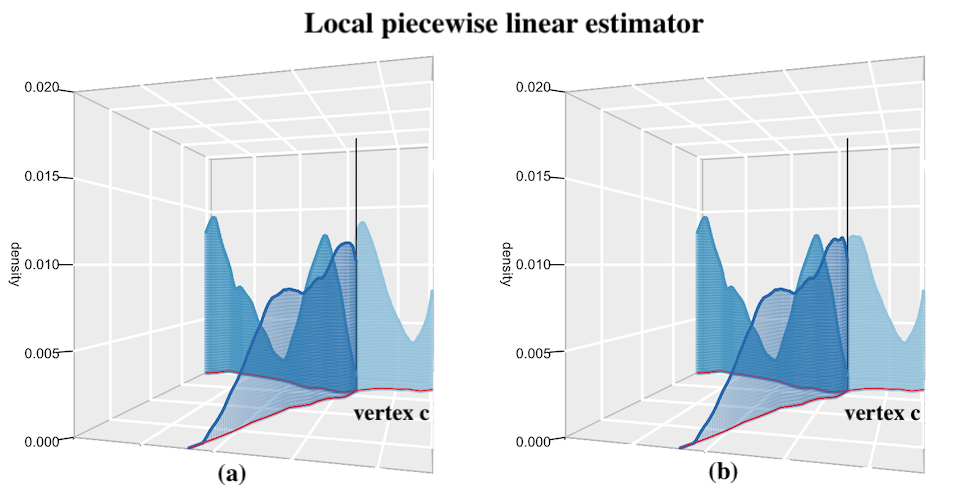}
\caption{(a): Density estimation near vertex C by local piecewise estimator using only the data on each edge. (b):  Density estimation near vertex C by local piecewise estimator using data from all three edges subject to the constraint that the densities on the two edges in the foreground are equal at the vertex. The network is in red and the estimates are in blue.}\label{est_c_new}
\end{figure}
For vertex D,  under piecewise local linear estimation, there's evidence in the data that the densities from the two left edges are equal when approaching the vertex. Figure \ref{est_d_old} shows the estimated densities near vertex D, by ESCK, ESDK and DE, while piecewise local linear estimation is in Figure \ref{est_d_new}. From this example we see that the network is embedded in $\mathbb{R}^3$. The vertex is on the right of the plot where labeled ``vertex D''. There is an overpass on the left of the plot, where the two edges do not intersect. 
\begin{figure}[h]
\centering
\includegraphics[scale=0.7]{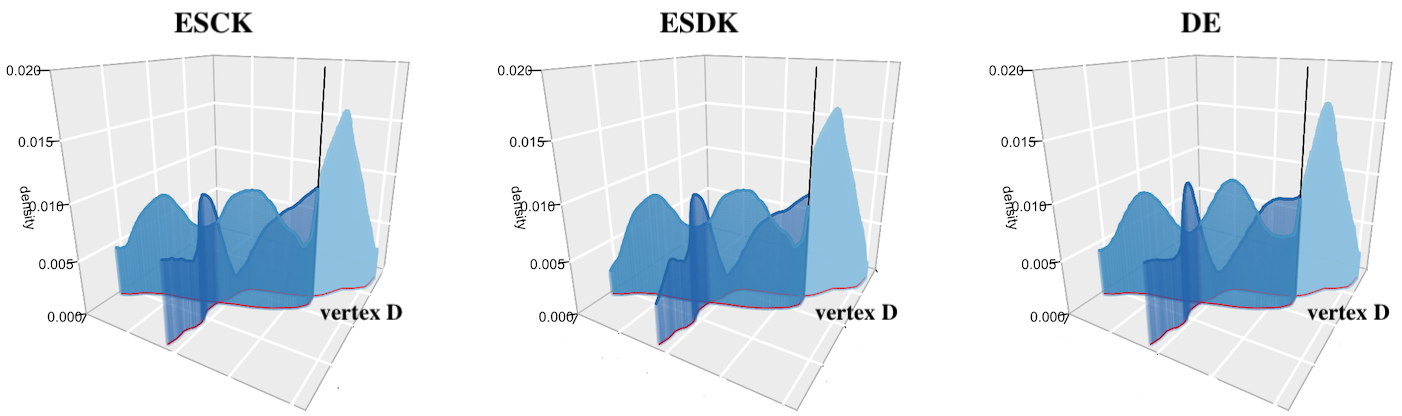}
\caption{Density estimation near vertex D, by ESCK, ESDK and DE. The network is in red and the estimates are in blue. The network is in red and the estimates are in blue.}\label{est_d_old}
\centering
\includegraphics[scale=0.8]{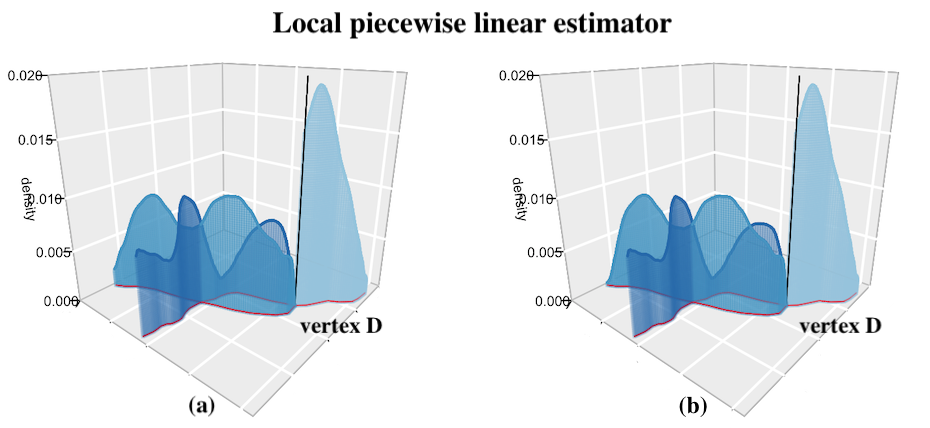}
\caption{(a): Density estimation near vertex D by local piecewise estimator using only the data on each edge. (b):  Density estimation near vertex D by local piecewise estimator using data from all three edges subject to the constraint that the densities on the left two edges are equal at the vertex. The network is in red and the estimates are in blue.}\label{est_d_new}
\end{figure}

\section{Discussion}
\label{sec:discuss}
As we mentioned in the introduction, there is great potential demand in many fields for estimating the density of events on a network. An easy method for such estimations is to use the ordinary kernel density estimation method that assumes an unbounded plane, or kernel density estimation on the real line with the Euclidean distance replaced by network distance. Many papers in the literature employ this method. However, this method yields a bias in density estimation and so the method is likely to lead to misleading conclusions. We also discussed the equal-split discontinuous kernel estimator, the equal-split continuous kernel estimator, and diffusion estimator. None of those methods allows for discontinuity in the estimates. The first two methods lack theoretical justification and are computationally expensive. The diffusion estimator is mathematically equivalent to an infinite-sum generalization of the equal-split continuous rule applied to the Gaussian density, and it inherits the asymptotic properties of a kernel density estimator \cite{mcswiggan2017kernel}. Also the diffusion estimator has a slower rate for the boundary bias.\\
\\
In this paper, we have formulated a density estimation procedure on a linear network via local piecewise polynomial regression by way of binning. We first apply local polynomial regression on each edge individually, then we test joint equality of the regression functions at the vertex. If the null hypothesis is not rejected, locations within the $h$-neighborhood of the vertex are re-estimated by local piecewise polynomial regression using data from all neighboring edges, subject to the equality constraint. The proposed piecewise linear procedure only imposes continuity at vertices when there is evidence in the data, and its asymptotic bias has the same rate at a vertex as an interior point. We studied the local linear case in detail, while there is a straightforward extension to higher-order polynomial approximation. When applying the proposed method to real data, if there are loops in the network, for simplicity, we only considered the shortest path between points, and we showed that only data points that have direct access to the evaluation point contribute to the estimation of the regression function.\\
\\
Due to space limitation, we have only considered only fixed-bandwidth smoothing, and we have not considered data-based bandwidth selection nor adaptive smoothing on a network. 
We proposed a test of equal intercepts at a vertex to decide whether to assume equal intercepts when estimating the density in the neighborhood of the vertex.  In the future, we will consider an estimator that, rather than using a pre-test, shrinks the unequal-intercepts estimator towards the equal-intercepts estimator. We have assumed that all locations are measured without error and lie exactly on the linear network. However this is not true for some applications, such as ambulance or taxi, where there are GPS error in their locations. Further study into such measurement error problems is required. In addition to estimating probability density or point process intensity, the proposed procedure is also used for regression problems, such as varying coefficient models on a linear network.

\section{Proofs}
\begin{proof}[Proof of Theorem 1]
By Taylor expansion,
\begin{align}
    E\left(\hat{m}(x)-m(x)\right)=\boldsymbol{e}_1^T\omega^{-1}\left(\boldsymbol{X}^T\boldsymbol{W}\boldsymbol{X}\right)^{-1}\omega\boldsymbol{X}^T\boldsymbol{W}\frac{1}{2}m^{(2)}(x)\begin{bmatrix}
    (x_1-x)^2  \\
    \vdots \\
    (x_n-x)^2
\end{bmatrix}+O\begin{bmatrix}
    (x_1-x)^3  \\
    \vdots \\
    (x_n-x)^3
\end{bmatrix}.
\end{align}
Note that if $m$ is a linear function then $m^{(r)}(x)=0$ for $r\ge2$ so that the local linear estimator is exactly unbiased when $m$ is a linear function. To find the leading bias term for general function $m$, note that
\begin{align}
\omega\boldsymbol{X}^T\boldsymbol{W}\boldsymbol{X}= \begin{bmatrix}
    \sigma^0_K       & h\sigma^1_K \\
    h\sigma^1_K       & h^2\sigma^2_K
\end{bmatrix}+O(\omega)\>\>{\rm and}\>\>\omega\boldsymbol{X}^T\boldsymbol{W}\begin{bmatrix}
    (x_1-x)^2  \\
    \vdots \\
    (x_n-x)^2
\end{bmatrix}=\begin{bmatrix}
    h^2\sigma^2_K  \\
h^3\sigma^3_K
\end{bmatrix}+O(\omega),
\end{align}
where $\sigma^i_K=\int u^iK(u)du$. Some straightforward matrix algebra then leads to the following expression for the leading bias term
\begin{align}
    E\left(\hat{m}(x)-m(x)\right)=\frac{1}{2}h^2\left[\frac{\left(\sigma^2_K\right)^2-\sigma^3_K\sigma^1_K}{\sigma^2_K-\left(\sigma^1_K\right)^2}\right]m^{(2)}(x)+O\left(\omega\right)+o\left(h^2\right).
\end{align}
To derive the asymptotic variance of $\hat{m}(x)$ we have
\begin{align}
        Var(\hat{m}(x))&=\omega\boldsymbol{e}^T_1\left(\omega\boldsymbol{X}^T\boldsymbol{W}\boldsymbol{X}\right)^{-1}\left(\omega\boldsymbol{X}^T\boldsymbol{W}\boldsymbol{V}\boldsymbol{W}\boldsymbol{X}\right)\left(\omega\boldsymbol{X}^T\boldsymbol{W}\boldsymbol{X}\right)^{-1}\boldsymbol{e}_1,
\end{align}
where $\boldsymbol{V}=diag(f(x_1),\dots,f(x_n))$, with $f(x_i)=(1/N\omega)m(x_i)-(1/N)m(x_i)^2$. We stress that $f(x_i)$ depends on $N$ and $\omega$, but we drop them from its notation for simplicity. Now use approximation analogous to those used above we have
\begin{align}
\omega\left(\boldsymbol{X}^T\boldsymbol{W}\boldsymbol{V}\boldsymbol{W}\boldsymbol{X}\right)=
\begin{bmatrix}
    \frac{1}{h}f(x)R^0_K+o\left(\frac{1}{h}\right)       & f(x)R^1_K+o(1) \\
    f(x)R^1_K+o(1)     & hf(x)R^2_K+o(h)
\end{bmatrix}+O(\omega),
\end{align}
where $R^i_K=\int u^iK(u)^2du$. These expressions can be combined to obtain
\begin{align}
    Var\left(\hat{m}(x)\right)=C(N,h,\omega,x)Q_K+o\left(\frac{1}{Nh}\right)+o\left(\frac{\omega}{Nh}\right),
\end{align}
where $C(N,h,\omega,x)=f(x)\omega/h$, and
\begin{align}
    Q_K=\frac{R^0_K\left(\sigma^2_K\right)^2-2R^1_K\sigma^2_K\sigma^1_K+R^2_K\left(\sigma^1_K\right)^2}{\left[\sigma^2_K-\left(\sigma^1_K\right)^2\right]^2}.
\end{align}
\end{proof}
%%%%%%%%%%%%%%%%%%%%%%%%%%%%%%%%%%%%%%%%%%%%%%%%%%%%%%%
%%%%%%%%%%%%%%%%%%%%%%%%%%%%%%%%%%%%%%%%%%%%%%%%%%%%%%%

\begin{proof}[Proof of Theorem 2]
Recall the asymptotic result of local polynomial regression:
\begin{align}
\sqrt{nh}\left[\hat{m}_{e_l}(x)-m_{e_l}(x)-B_{e_l}\right]\xrightarrow[]{d}N\left(0,V_{e_l}\right),\notag
\end{align}
where $B_{e_l}$ and $V_{e_l}$ are the asymptotic bias and variance of $\hat{m}_{e_l}(x)$. Let $\hat{\boldsymbol{m}}=(\hat{m}_{e_1}(v),\dots, \hat{m}_{e_J}(v))^T$, we have asymptotically, $h\to0$ and $nh\to\infty$,
\begin{align}
\hat{\boldsymbol{m}}
\stackrel{a}{\sim}
N\left(\begin{bmatrix}
    {m}_{e_1}(x) +B_{e_1} \\
    \vdots \\
    {m}_{e_J}(x)+B_{e_J}
\end{bmatrix},
\begin{bmatrix}
    V_{e_1}        & \dots & 0 \\
    \vdots & \ddots & \vdots\\
   0        &  \dots & V_{e_J}
\end{bmatrix}
\right).\notag
\end{align}
Now consider a $(J-1)\times J$ contrast matrix $\boldsymbol{C}$ such that $\boldsymbol{C1}=\boldsymbol{0}$ (i.e. each row sum to zero). One choice of $\boldsymbol{C}$ is 
\begin{align}
\boldsymbol{C}=
\begin{bmatrix} 1 & -1 & 0 & 0 & \dots \\ 1 & 0& -1 & 0 & \dots \\ 1 & 0 & 0 & -1 & \dots \\\vdots & \vdots & \vdots & \vdots & \vdots &\\ \end{bmatrix},\notag
\end{align}
which simply contrasts edge 1 with edge 2, edge 1 with edge 3. Then
\begin{align}
\boldsymbol{C}\hat{\boldsymbol{m}}\stackrel{a}{\sim} 
N\left(\boldsymbol{\mu},\boldsymbol{\Sigma}\right)\notag
\end{align}
where
\begin{align}
\boldsymbol{\mu}=
\begin{bmatrix}
    {m}_{e_1}(x)-{m}_{e_2}(x)+B_{e_1}-B_{e_2} \\
    \vdots \\
    {m}_{e_1}(x)-{m}_{e_J}(x)+B_{e_1}-B_{e_J}
\end{bmatrix}\>\>{\rm and}\>\>\boldsymbol{\Sigma}=
\begin{bmatrix}
    V_{e_1}+V_{e_2}      & \dots & V_{e_1} \\
    \vdots & \ddots & \vdots\\
   V_{e_1}   &  \dots & V_{e_1}+V_{e_J}
\end{bmatrix}.\notag
\end{align}
Under the null hypothesis, 
\begin{align}
\boldsymbol{C}\hat{\boldsymbol{m}}
\stackrel{a}{\sim} N\left(\boldsymbol{0}, \boldsymbol{\Sigma}\right).\notag
\end{align}
We want to calculate the probability of generating a point at least as unlikely as the observed data point. To do that, we note that under the null hypothesis, the Mahalanobis distance follows a chi-square distribution 
\begin{align}
\left(\boldsymbol{C}\hat{\boldsymbol{m}}\right)^T\boldsymbol{\Sigma}^{-1}\boldsymbol{C}\hat{\boldsymbol{m}}\stackrel{a}{\sim} \chi^2_{J-1}.\notag
\end{align}
The test statistic is invariant under the choice of contrast matrices. This is easy to see by noticing that the rows of $\boldsymbol{C}$ are linearly independent. So we have a basis of some vector space $\mathbf{V}$ (and it doesn't matter if $\mathbf{V}$ is all of $\mathbb{R}^{p+1}$, or some subspace thereof), and two different ordered bases for $\mathbf{V}$, $\boldsymbol{b}_1$ and $\boldsymbol{b}_2$ (necessarily of the same size, since two bases of the same vector space always have the same size, here they are the transpose of two contrast matrices $\boldsymbol{C}_1$ and $\boldsymbol{C}_2$):
\begin{align*}
\boldsymbol{b}_1 &= \Bigl[ \mathbf{v}_1,\mathbf{v}_2,\ldots,\mathbf{v}_n\Bigr]\\\
\boldsymbol{b}_2 &= \Bigl[ \mathbf{w}_1,\mathbf{w}_2,\ldots,\mathbf{w}_n\Bigr].
\end{align*}
A change-of-basis matrix is a matrix that translates from $\boldsymbol{b}_1$ coordinates to $\boldsymbol{b}_2$ coordinates. That is, $A$ is a change-of-basis matrix (from $\boldsymbol{b}_1$ to $\boldsymbol{b}_2$) if, given the coordinate vector $[\mathbf{x}]_{\boldsymbol{b}_1}$ of a vector $\mathbf{x}$ relative to $\boldsymbol{b}_1$, then $A[\mathbf{x}]_{\boldsymbol{b}_1}=[\mathbf{x}]_{\boldsymbol{b}_2}$ gives the coordinate vector of $\mathbf{x}$ relative to $\boldsymbol{b}_2$, for all $\mathbf{x}$ in $\mathbf{V}$.\\
\\
To get a change-of-basis matrix, we write each vector of $\boldsymbol{b}_1$ in terms of $\boldsymbol{b}_2$, and these are the columns of $A$, for $i=1,\dots,n$, $\mathbf{v}_i = a_{1i}\mathbf{w}_1 + a_{2i}\mathbf{w}_2 + \cdots + a_{ni}\mathbf{w}_n$. We know we can do this because $\boldsymbol{b}_2$ is a basis, so we can express any vector (in particular, the vectors in $\boldsymbol{b}_1$) as linear combinations of the vectors in $\boldsymbol{b}_2$. Then the change-of-basis matrix translating from $\boldsymbol{b}_1$ to $\boldsymbol{b}_2$ is
\begin{align}
A=
\begin{bmatrix}
a_{11}  & \cdots & a_{1n}\\
\vdots  & \ddots & \vdots \\
a_{n1} & \cdots & a_{nn}
\end{bmatrix}.\notag
\end{align}
Matrix $A$ is always invertible. This is because just like there is a change-of-basis from $\boldsymbol{b}_1$ to $\boldsymbol{b}_2$, there is also a change-of-basis from $\boldsymbol{b}_2$ to $\boldsymbol{b}_1$. Since $\boldsymbol{b}_1$ is a basis, we can express every vector in $\boldsymbol{b}_2$ using the vectors in $\boldsymbol{b}_1$, for $i=1,\dots,n$, $\mathbf{w}_i = b_{1i}\mathbf{v}_1 + b_{2i}\mathbf{v}_2 + \cdots + b_{ni}\mathbf{v}_n$. So the matrix $B$, with
\begin{align}
B=
\begin{bmatrix}
b_{11}  & \cdots & b_{1n}\\
\vdots  & \ddots & \vdots \\
b_{n1}  & \cdots & b_{nn}
\end{bmatrix},\notag
\end{align}
has the property that given any vector $\mathbf{x}$, if $[\mathbf{x}]_{\boldsymbol{b}_2}$ is the coordinate vector of $\mathbf{x}$ relative to $\boldsymbol{b}_2$, then $B[\mathbf{x}]_{\boldsymbol{b}_2}=[\mathbf{x}]_{\boldsymbol{b}_1}$ is the coordinate vector of $\mathbf{x}$ relative to $\boldsymbol{b}_1$. Then applying first $A$ and then $B$ translates $\boldsymbol{b}_1$ coordinates into $\boldsymbol{b}_2$ coordinates and back to $\boldsymbol{b}_1$ coordinates, and thus $AB$ must be the identity matrix (likewise for $BA$). So $A$ and $B$ are both invertible, and every change-of-basis matrix is necessarily invertible.
\end{proof}
%%%%%%%%%%%%%%%%%%%%%%%%%%%%%%%%%%%%%%%%%%%%%%%%%%%%%%%%%%%%%%%%%%%%%%%%%%%%%%%%%%%%%%%%%%%%%%%%%%%%%%%%%%%%%%%%%%%%%%%%%%%%%%%%%%%%%%%%%%%%%%%%%%%%%%%%%%%%%%%%%%%%%%%%%%%%%%%%%%%%%%%%%%%%%%%%%%%%%%%%%%%%%%%%%%%%%%%%%%%%%%%%%%%%%%%%%%%%%%%%%%%%%%%%%%%%%%%%%%%%%%%%%%%%%%%%%%%%%%%%%%%%%%%%%%%%%%%%%%%%%%%%%%%%%%%%%%%%%%%%%%%%%%%%%%%%%%%%%%%%%%%%%%%%%%%%%%%%%%%%%%%%%%%%

\begin{proof}[Proof of Theorem 3]
Standard calculation shows that the leading term of the bias is given by
\begin{align}
    \frac{1}{2}\boldsymbol{e}_1^T\omega^{-1}\left(\boldsymbol{X}^T\boldsymbol{W}\boldsymbol{X}\right)^{-1}\omega\boldsymbol{X}^T\boldsymbol{W}\boldsymbol{S},\notag
\end{align}
where $\boldsymbol{S}$ is a column vector with entries being $(v-x)^2m^{(2)}_{e_l}(x)+(x_{ij}-v)^2m^{(2)}_{e_i}(v)$, for $i=1,\dots,n_j$, and $j=1,\dots,J$. Note that for data points $x_{ij}\in e_l$, where $e_l$ is the edge that the evaluation point $x$ is located, the entries are simplified to $(x_{il}-x)^2m^{(2)}_{e_l}(x)$, for $i=1,\dots,n_l$. To approximate $\omega^{-1}\left(\boldsymbol{X}^T\boldsymbol{W}\boldsymbol{X}\right)^{-1}$, we note that
\begin{align}
    &\omega\sum_i(x_{ij}-v)K_h(x_{ij}-x)=h\mu^{(j)}_1-(v-x)\mu^{(j)}_0+O\left(\omega\right),\notag\\
    &\omega\sum_i(x_{ij}-v)^2K_h(x_{ij}-x)=h^2\mu^{(j)}_2+(v-x)^2\mu^{(j)}_0-2h(v-x)\mu^{(j)}_1-O\left((v-x)\omega\right)+O\left(\omega\right),\notag\\
&\omega\sum_i(x_{il}-x)^pK_h(x_{il}-x)=h^p\mu^{(l)}_p+O\left(\omega\right)\>\>{\rm for}\>\>p=0,1,2,\notag
\end{align}
where $\mu^{(j)}_i=\int_{e_j}u^iK(u)du$. If follows that
\begin{align}
    \omega\boldsymbol{X}^T\boldsymbol{W}\boldsymbol{X}=\boldsymbol{A}\left[\boldsymbol{U}_0+\left(\frac{v-x}{h}\right)\boldsymbol{U}_1+\left(\frac{v-x}{h}\right)^2\boldsymbol{U}_2\right]\boldsymbol{A}+O(\omega\boldsymbol{1}),\notag
\end{align}
where $\boldsymbol{A}=diag(1,h,\dots,h)$. The matrix $\boldsymbol{U}_0$ is symmetric, and its first row is $\sum_j\mu^{(j)}_0,\mu^{(1)}_1,\dots,\mu^{(J)}_1$ and its diagonal is $\sum_j\mu^{(j)}_0,\mu^{(1)}_2,\dots,\mu^{(J)}_2$. All other entries are zero. We also have that matrix $\boldsymbol{U}_1$ is symmetric, and its first row is $0,-\mu^{(1)}_0,\cdots,-\mu^{(l-1)}_0,\sum_{j\ne l}\mu^{(j)}_0,-\mu^{(l+1)}_0,\cdots,-\mu^{(J)}_0$. Its $(l+1)$th row is $\sum_{j\ne l}\mu^{(j)}_0,\mu^{(1)}_1,\cdots,\mu^{(l-1)}_1,0,\mu^{(l+1)}_1,\cdots,\mu^{(J)}_1$ and its diagonal is $0,-2\mu^{(1)}_1,\cdots,-2\mu^{(l-1)}_1,$ $0,-2\mu^{(l+1)}_1,\cdots,-2\mu^{(J)}_1$. All other entries of $\boldsymbol{U}_1$ are zero. Finally $\boldsymbol{U}_2=diag(0,\mu^{(1)}_0,\dots,\mu^{(J)}_0)$. Take inverse we have
\begin{align}
    \omega^{-1}\left(\boldsymbol{X}^T\boldsymbol{W}\boldsymbol{X}\right)^{-1}
    &=\boldsymbol{A}^{-1}\left[\boldsymbol{U}_0^{-1}-\left(\frac{v-x}{h}\right)\boldsymbol{U}_0^{-1}\boldsymbol{U}_1\boldsymbol{U}_0^{-1}+\left(\frac{v-x}{h}\right)^2\boldsymbol{U}_0^{-1}\boldsymbol{U}_2\boldsymbol{U}_0^{-1}\right]\boldsymbol{A}^{-1}\notag\\
    &+o\left(\left(\frac{v-x}{h}\right)^2\boldsymbol{A}^{-1}\boldsymbol{1}\boldsymbol{A}^{-1}\right)+O(\omega\boldsymbol{A}^{-1}\boldsymbol{1}\boldsymbol{A}^{-1}).\notag
\end{align}
To approximate $\omega\boldsymbol{X}^T\boldsymbol{W}\boldsymbol{S}$, we note that
\begin{align}
\sum_i\left(x_{ij}-v\right)^3m^{(2)}_{e_j}(v)K_h(x_{ij}-x)=&h^3\mu^{(j)}_3-(v-x)^3\mu^{(j)}_0-3(v-x)h^2\mu^{(j)}_2+3(v-x)^2h\mu^{(j)}_1\notag\\
&+O((v-x)\omega)+O((v-x)^2\omega)+O((v-x)^3\omega)+O(\omega).\notag
\end{align}
Combining this result with the approximations before, we get
\begin{align}
    \omega\boldsymbol{X}^T\boldsymbol{W}\boldsymbol{S}=\boldsymbol{A}\left[h^2\boldsymbol{R}_0+h(v-x)\boldsymbol{R}_1+(v-x)^2\boldsymbol{R}_2+\frac{(v-x)^3}{h}\boldsymbol{R}_3\right]+O(\omega),\notag
\end{align}
where $\boldsymbol{R}_0$, $\boldsymbol{R}_1$, $\boldsymbol{R}_2$ and $\boldsymbol{R}_3$ are column vectors. The first entry of $\boldsymbol{R}_0$ is $\sum_{j\ne l}m^{(2)}_{e_j}(v)\mu^{(j)}_2+m^{(2)}_{e_l}(x)\mu^{(l)}_2$, and the rest are $m^{(2)}_{e_j}(v)\mu^{(j)}_3$, for $j=1,\cdots,J$. The first entry of $\boldsymbol{R}_1$ is $-2\sum_{j\ne l}m^{(2)}_{e_j}(v)\mu^{(j)}_1$, and the rest, except the $(l+1)$th entry, are $-3m^{(2)}_{e_j}(v)\mu^{(j)}_2$, for $j\ne l+1$, and the $(l+1)$th entry is $\sum_{j\ne l}m^{(2)}_{e_{j}}(x)\mu^{(j)}_2$. The first entry of $\boldsymbol{R}_2$ is $m^{(2)}_{e_l}(x)\sum_{j\ne l}\mu^{(j)}_0+\sum_{j\ne l}m^{(2)}_{e_j}(x)\mu^{(j)}_0$, and the rest, except the $(l+1)$th entry, are $m^{(2)}_{e_l}(x)\mu^{(j)}_1+3m^{(2)}_{e_j}(v)\mu^{(j)}_1$, for $j\ne l+1$, and the $(l+1)$th entry is $-2\sum_{j\ne l}m^{(2)}_{e_j}(v)\mu^{(j)}_1$. Finally, the first entry of $\boldsymbol{R}_3$ is $0$, and the rest, except the $(l+1)$th entry, are $-m^{(2)}_{e_l}(x)\mu^{(j)}_0-m^{(2)}_{e_j}(v)\mu^{(j)}_0$, for $j\ne l+1$, and the $(l+1)$th entry is $m^{(2)}_{e_l}(x)\sum_{j\ne l}\mu^{(j)}_0+\sum_{j\ne l}m^{(2)}_{e_j}(v)\mu^{(j)}_0$. Consequently,
\begin{align}
        E\left(\hat{m}(x)-m(x)\right)
        =\frac{1}{2}\boldsymbol{e}^T_1\boldsymbol{U}_0^{-1}\left(h^2\boldsymbol{R}_0+h(v-x)\left(\boldsymbol{R}_1-\boldsymbol{U}_1\boldsymbol{U}_0^{-1}\boldsymbol{R}_0\right)\right)+o\left(h(v-x)\right)+o\left(h^2\right)+O(\omega).\notag
\end{align}
To derive the asymptotic variance of $\hat{m}(x)$ we have
\begin{align}
        Var(\hat{m}(x))&=\boldsymbol{e}^T_1\left(\boldsymbol{X}^T\boldsymbol{W}\boldsymbol{X}\right)^{-1}\left(\boldsymbol{X}^T\boldsymbol{W}\boldsymbol{V}\boldsymbol{W}\boldsymbol{X}\right)\left(\boldsymbol{X}^T\boldsymbol{W}\boldsymbol{X}\right)^{-1}\boldsymbol{e}_1,\notag
\end{align}
where $\boldsymbol{V}=diag(f(x_1),\dots,f(x_n))$, with $f(x_i)=(1/N\omega)m(x_i)-(1/N)m(x_i)^2$. We stress that $f(x_i)$ depends on $N$ and $\omega$, but we drop them from its notation for simplicity. Now use approximation analogous to those used above we have
\begin{align}
    &\omega\sum_i(x_{ij}-v)K_h(x_{ij}-x)^2f(x_{ij})=f(x)R^{(j)}_1-\left(\frac{v-x}{h}\right)f(x)R^{(j)}_0-o\left(\frac{v-x}{h}\right)+O\left(\omega\right),\notag\\
    &\omega\sum_i(x_{ij}-v)^2K_h(x_{ij}-x)^2f(x_{ij})\notag\\
    &=hf(x)R^{(j)}_2-2(v-x)f(x)R^{(j)}_1+\left(\frac{v-x}{h}\right)^2f(x)R^{(j)}_0+o(h)+o\left(\frac{(v-x)^2}{h}\right)+O\left(\omega\right),\notag\\
    &\omega\sum_i(x_{il}-x)^pK_h(x_{il}-x)^2f(x_{ij})=h^{p-1}f(x)R^{(j)}_p+o(h^{p-1})+O\left(\omega\right),\>\>{\rm for}\>\>p=0,1,2.\notag
\end{align}
It follows that
\begin{align}
    &\omega\boldsymbol{X}^T\boldsymbol{W}\boldsymbol{V}\boldsymbol{W}\boldsymbol{X}\notag\\
    &=\frac{1}{h}f(x)\boldsymbol{A}\left[\boldsymbol{M}_0+\left(\frac{v-x}{h}\right)\boldsymbol{M}_1+\left(\frac{v-x}{h}\right)^2\boldsymbol{M}_2+o\left(\frac{v-x}{h}\right)++o\left(\left(\frac{v-x}{h}\right)^2\right)\right]\boldsymbol{A},\notag
\end{align}
where $\boldsymbol{M}_0$ is symmetric, its first row is $\sum_jR_0^{(j)}, R_1^{(1)},\cdots,R_1^{(J)}$, and its diagonal entries are $\sum_jR_0^{(j)}, R_2^{(1)},\cdots,R_2^{(J)}$. All the other entries of $\boldsymbol{M}_0$ are zero. $\boldsymbol{M}_1$ is symmetric, its first row is $0,-R^{(1)}_0,\cdots,-R^{(l-1)}_0,\sum_jR^{(j)}_0,-R^{(l+1)}_0,\cdots,-R^{(J)}_0$, its $(l+1)$th row is $\sum_jR^{(j)}_0,R^{(1)}_1,\cdots,R^{(l-1)}_1,0,$ $R^{(l+1)}_1,\cdots,R^{(J)}_1$, and its diagonal is $0,-R^{(1)}_1,\cdots,-R^{(l-1)}_1,0,-R^{(l+1)}_1,\cdots,-R^{(J)}_1$. All other entries of $\boldsymbol{M}_1$ are zero. Finally $\boldsymbol{M}_2$ is also symmetric, its first row is the zero vector, its $(l+1)$th row is $0,-R^{(0)}_1,\cdots,-R^{(l-1)}_1,\sum_jR^{(j)}_0,-R^{(l+1)}_1,\cdots,-R^{(J)}_1$, and its diagonal is $0, R^{(1)}_0,\cdots,R^{(l-1)}_0,\sum_jR^{(j)}_0,$ $R^{(l+1)}_0,\cdots,R^{(J)}_0$. All other entries of $\boldsymbol{M}_2$ are zero.\\
\\
Combining the above expressions, we have
\begin{align}
    Var(\hat{m}(x))&=C(N,h,\omega,x)\left[\boldsymbol{U}^{-1}_0\boldsymbol{M}_0\boldsymbol{U}^{-1}_0+\left(\frac{v-x}{h}\right)\boldsymbol{U}^{-1}_0\left(\boldsymbol{M}_0\boldsymbol{U}^{-1}_0\boldsymbol{U}_1+\boldsymbol{M}_1-\boldsymbol{U}_1\boldsymbol{U}^{-1}_0\boldsymbol{M}_0\right)\boldsymbol{U}^{-1}_0\right]\notag\\
    &+o\left(\frac{1}{Nh}\right)+o\left(\frac{\omega}{Nh}\right)+o\left(\frac{v-x}{Nh^2}\right)+o\left(\frac{\omega(v-x)}{Nh^2}\right).
\end{align}
\end{proof}

\section*{Supplementary Materials}

The supplementary materials include an R program containing code to perform the local linear regression method for density estimation on a network as described in this article. The program also contains all codes for simulating datasets used in the article.

\end{document}